# Formation energies of charged defects in two-dimensional materials – resolution of long-standing difficulties

Andrew O'Hara[1], Blair R. Tuttle[1,2], Xiao-Guang Zhang[3], and Sokrates T. Pantelides[1*]

[1]*Department of Physics and Astronomy and Department of Electrical Engineering and Computer Science, Vanderbilt University, Nashville, TN, 37235*

[2]*Department of Physics, Penn State Behrend, Erie, PA 16563*

[3]*Department of Physics and the Quantum Theory Project, University of Florida, Gainesville, FL 32611*

## Abstract

Formation energies of charged point defects in semiconductors are calculated using periodic supercells, which entail a divergence arising from long-range Coulombic interactions. The divergence is typically removed by the so-called jellium approach. Recently, Wu, Zhang and Pantelides [WZP, *Phys. Rev. Lett*. **119** 105501 (2017)] traced the origin of the divergence to the assumption that charged defects are formed by physically removing electrons from or adding electrons to the crystal, violating charge neutrality, a key principle of statistical mechanics that determines the Fermi level. An alternative theory was constructed by recognizing that "charged" defects form by trading carriers with the energy bands, whereby supercells are always charge-neutral so that *no divergence is present and no ad-hoc procedures need to be adopted for calculations*. Here we give a more detailed exposition of the foundations of both methods and show that the jellium approach can be derived from the statistical-mechanics-backed WZP definition by steps whose validity cannot be assessed *a priori*. In particular, the divergence appears when the charge density of band carriers is dropped, leaving a supercharged crystal. In the case of charged defects in two-dimensional (2D) materials, unphysical fields appear in vacuum regions. None of these pathological features are present in the reformulated theory. Finally, we report new calculations in both bulk and 2D materials. The WZP approach yields formation energies that differ from jellium values by up to ~1 eV. By analyzing the spatial distribution of wave functions and defect potentials, we provide insights into the inner workings of both methods and demonstrate that the failure of the jellium approach to include the neutralizing electron density of band carriers, as is the case in the physical system, is responsible for the numerical differences between the two methods.

*pantelides@vanderbilt.edu

I. **Introduction**

The formation energy of a point defect in a crystal determines the concentration of the defect under equilibrium conditions.[1, 2] Differences in the formation energies of charge states yield the defect's thermodynamic (charge-transition) energy levels in the band gap, which play a key role in using electrical measurements to identify the defects that are responsible for device degradation.[3] In addition to bulk crystals, defects and their formation energies are important at surfaces, e.g., as catalyst active sites.[4] The recent explosion of interest in two-dimensional (2D) materials has brought a major focus on defects that are either a detriment or a way to induce desired functionalities.[5] Once more, defect formation energies play a central role. In particular, the performance of monolayer materials in electrical devices depends critically on the control of charged defects, which act as scattering centers and recombination sites.

Calculations of formation energies of defects in semiconductors using density functional theory (DFT) were first carried out in the 1980's using Green's functions, which effectively yield the energy cost to introduce a single defect in an otherwise perfect infinite solid.[6-8] When a defect entails a change in the number of individual atomic species that are present, e.g., a vacancy, an intersitial, or an impurity, a corresponding chemical potential needs to be inroduced to account for the accompanying change in energy. The same approach was adopted for charged defects, adding the energy gain or cost for electrons to be removed to or taken from a hypothetical state at the Fermi level, which represents the electron chemical potential.

Periodic supercells containing a single defect were first used for calculations of defect energy levels in 1976.[9] It was only possible to use supercells containing 54 atoms, which were found to be too small, resulting in large dispersions of bound states in the energy gap. Supercells for total-energy calculations (migration barriers) were first used with very small supercells, 8 and 16 atoms,[10] but as computer performance improved, supercell calculations for formation energies became the method of choice. One could in principle employ successively larger, periodically repeated supercells until the formation energy converges, i.e., it reaches a plateau as a function of supercell size. If convergence to a value is not reached, one can then extrapolate the finite-supercell values to an infinite supercell. Either way, the infinite-supercell limit should in principle match the value one would obtain using the Green's-function approach.

The supercell scheme works very well for neutral defects. In the case of charged defects, however, the standard definition of formation energy places a charged defect with a long-range Coulomb potential in every supercell, which leads to a divergence – the average electrostatic potential, namely the $G = 0$ component of the potential in reciprocal space, is infinite ($G$ are the reciprocal lattice vectors of the adopted supercell periodic structure). In the earliest total-energy supercell calculations for charged defects,[10] the divergence was removed by setting the $G = 0$ component of the potential to zero. This step was interpreted to mean that an "inert" uniform background charge is used to neutralize the system on the grounds that such a charge affects only the $G = 0$ of the potential. The name "jellium" was later used to refer to this procedure,[2] borrowing the term from its usage to describe a homogeneous electron gas as an approximation to the electrons in a simple metal. This homogeneous background charge density, however, is *not* included explicitly in the calculation of the total charge density in the self-consistency loop.

Nevertheless, "setting the average electrostatic potential to zero" or "assuming the presence of a compesnating uniform background charge density" are generally used interchangeably in the literature as a process that neuralizes the supercells. Alternative procedures have also been invoked to remove the divergence.[11-14]

For many years, for computational reasons, charged-defect jellium calculations were routinely performed for the largest possible supercells, typically fewer than 100 atoms by the early 2000's[2] (careful calculations for the *neutral* vacancy in Si had, however, shown in 1998 that, when only the Γ point is used to sample the Brillouin zone, it takes 216-atom supercells to obtain converged results[15]). As far back as 1985, however, it was recognized by Leslie and Gillan[16] that, in the jellium approach, charged-defect formation energies do not converge as the supercell size is increased, i.e., they do not reach a plateau as is generally the case for neutral defects. Though setting $V(G = 0) = 0$ eliminates the divergence, it does not remove the long-range Coulomb interactions between the defect's periodic images. It became obvious that it is necessary to interpolate the calculated values to the inifnite-supercell limit. In this sense, the long-range interactions were viewed as artifacts of finite-supercell calculations. Leslie and Gillan[16] recognized that the calculated formation energies for charged defects scale roughly as $L^{-1}$, where $L$ is the linear dimension of the supercell. Furthermore, they showed that this scaling arises from the fact that, for relatively large supercells, the long-range interactions have the same form as those of a periodic array of point charges emnedded in a uniform background charge density of opposite sign, a form of Madelung energy, which indeed scales as $L^{-1}$. It is then possible to subtract such a model contribution from the DFT-calculated formation energies and obtain *a-posteriori* "corrected" values, which converge quickly and even justify a single-supercell calculation as adequate.

The seminal work of Leslie and Gillam[16] inspired more sophisticated schemes by Makov and Payne[17] and additional refinements[18-25] to produce "corrected" values that in principle converge quickly to the extrapolated value of the DFT-calculated "uncorrected" formation energies. In 2008, Lany and Zunger[19] pointed out that just the Makov-Payne correction term that scales as $L^{-1}$ works just fine to remove the long-range interactions if other finite-size effects are carefully eliminated. An insightful commentary on the many developments on finite-size corrections was given in 2009 by Nieminen.[26] A comprehensive comparison of several "finite-supercell correction" schemes was given in 2012 by Komsa, Rantala, and Pasquarello.[27] They showed that the "corrected" formation energies for several defects converge to the same value as the extrapolated "uncorrected" DFT-calculated formation energies, with the scheme developed by Freysoldt, Neugebaur, and van de Walle in 2011[23] providing the best results. In some notable cases, however, the "corrected" and "uncorrected' values do not extrapolate to the same infinite-supercell limit for reasons that were understood – the conditions that are necessry for the validity of the correction schemes are not fully satisfied. Clearly, an element of uncertainty is present in relying fully on small supercells and correction schemes.

The effects of artificial long-range interactions between images of low-dimensional systems such as molecules, nanoparticles, nanotubes, and slabs, which require a vacuum region in periodically-repeated supercells, have been an issue for general electronic-structure calculations. Their avoidance has been addressed by several authors in both real-space and reciprocal-space calculations since the early 1990's,[28-32] and were reviewed by Dabo et al. in 2008.[33]. However, the artificial long-range interactions that are encountered in supercell calculations of charged-defect formation energies at surfaces and in 2D materials, which also require vacuum regions in the supercells, present unique challenges.

Charged defects at surfaces were first studied using the jellium scheme in the 1990's.[34] In 2013, several authors[13, 14, 35] showed that, in the jellium scheme, the formation energy of a charge defect diverges linearly with the size of the vacuum region if the in-plane supercell dimensions are held fixed. This additional divergence was attributed to the jellium scheme's effective uniform background charge in the vacuum region, which sets up a diverging dipole.[14] Komsa and Pasquarello[35] retained the jellium model and showed that the divergence is not present if all three supercell dimensions are scaled uniformly. Richter et al.[13] and Moll et al.[14], resolved the unphysical divergence by abandoning the jellium scheme. Instead of setting $V(\boldsymbol{G}=0)=0$, they introduced explicitly a discretized neutralizing charge in the confines of the material, as in the virtual-crystal approximation (VCA). For example, for a positively charged defect, a compensating charge equal to $-e/N$ is placed on each nucleus ($e$ is the magnitude of the charge of an electron and $N$ is the number of nuclei in the supercell – all-electron calculations are used, which makes this procedure feasible). The method then can handle a charged defect at a surface in the same manner as a charged defect in a bulk material.

Komsa and Pasquarello[35] adapted the *a posteriori* correction scheme of Freysoldt et al.[23] for charged defects at surfaces by dealing with a new complication -- the construction of a macroscopic spatially dependent dielectric constant with sufficient atomic-scale spatial resolution to transition from unity in the vacuum region to a finite value in the bulk of the film. This dielectric constant is used to screen the model long-range interactions that are subtracted from the DFT-calculated formation-energy values. The long-range interactions through the vacuum regions of course remain unscreened, whereby the transition region features a prominent role in the calculation. Komsa and Pasquarello were sucessful in demonstrating that the "corrected" and "uncorrected" formation energies extrapolate to the same value at the infinite-supercell limit.

The case of charged defects at crystalline surfaces served as a foundation to address the case of charged defects in 2D materials, but new complications arise in implementing *a posteriori* finite-cell corrections.[36-39] In order to determine a suitable macroscopic-dielectric-constant profile, the "thickness" of a monolayer needs to be specified, which adds additional complications to the inherent uncertainties of calculating dielectric constants, e.g., arising from inaccurate band gaps. In 2014, Noh et al.[37] discovered yet another complication: Calculations of the *a posteriori* finite-cell correction for very large supercells revealed that, at some large value of the supercell dimension, the correction values acquire a curvature that requires a fifth-order polynomial to obtain a fit for extrapolation to the infinite-supercell limit. Since it was not practical to perform DFT calculations for very large supercells, *no test could be carried out to determine if the DFT-calculated values and the a posteriori-corrected values extrapolate to the same infinite-supercell-limit value* (recall that, as emphasized by Komsa et al.,[27] the extrapolated value of the uncorrected DFT-calculated values set the standard for judging the validity of the *a posteriori* corrections). In 2018, in an Erratum to their 2014 paper, Komsa et al.[39] confirmed that the large-supercell behavior of the model corrections deviates significantly from a small-order polynomial. The deviation was attributed to the inhomogeneity of the dielectric constant. To deduce infinite-supercell-limit formation energies, both Noh et al.[37] and Komsa et al.[39] assumed that the calculated formation energies have a similar extrapolation behavior. The net result is that, for charged-defects in 2D materials, the ability of *a posteriori* correction schemes to produce "corrected" formation energies that extrapolate to the same infinite-supercell-value as uncorrected DFT-calculated formation energies remains untested. More recent papers[40, 41] have focused on improving other aspects of the correction schemes, especially relating to the dielectric-constant models, but did not address this open issue.

We can summarize the state of the theory of charged-defect formation energies as follows. The cornerstone of the theory is the calculation of the energy cost to introduce an isolated charged defect in an otherwise perfect infinite crystal, which defines the dilute limit. The charged defect is generated from a neutral defect by removing from or adding to the crystal one or more electrons, while the energy gain or cost of removing or adding electrons is included relative to a hypothetical reservoir at the Fermi level, which defines the electron chemical potential. In a supercell approach, the long-range interactions among charged defects result in a divergent electrostatic potential. The divergence is viewed as an artifact of finite-size supercells and is removed, typically by the jellium approach, with the correct formation energy obtained by an extrapolation of finite-supercell DFT-calculated values to infinite-size supercell. To enable a reliable extrapolation using relatively small supercells or even employ just one supercell size, the contributions of the long-range interactions are removed by an *a posteriori* approach. For bulk charged defects in bulk semiconductors, tests have shown that the "corrected" formation energies typically, but not always, extrapolate to the same value as the extrapolation of "uncorrected" DFT-calculated formation energies, as they must.[27] For 2D materials, the *a posteriori* corrections in very large supercells deviate substantially from small-order polynomial fits and the ability of "corrected" formation energies to extrapolate to the same inifinite-superrcell-limit as the "uncorrected" DFT-calculated formation energies has not been tested. Though improved correction schemes continue to be developed, the extrapolation issue has remained open.

In a recent paper, Wu, Zhang and Pantelides (WZP)[42] revisited the definition of a charged defect's formation energy at a fundamental level and identified the origins of the various complications that we discussed so far, namely the divergences, the long-range Coulomb interactions that require extrapolations of calculated formation energies to the infinite-supercell limit, etc. It was noted that the conventional definition of charged-defect formation energies had never been derived from fundamental principles, but was merely an adaptation of the definition of the statistical-mechanics-backed definition of the formation energy of a neutral defect. To treat neutral defects such as vacancies, interstitials, or impurities, one removes atoms from or adds atoms to the crystal, with the energy cost or gain measured relative to the pertinent chemical potentials. The adaptation of the formation-energy definition to charged defects correctly retains this part for atom removals, additions, or exchanges, but fails to treat the conversion of neutral defects to "charged defects" in full accord with physical reality and the statistical mechanics of electrons in semiconductors: it handles the accounting of the energy cost or gain to remove electrons from or add electrons to a defect correctly relative to the Fermi level, which serves as the electron chemical potential, *but* *assumes that the electrons are actually removed from the crystal or added to the crystal, leaving behind truly charged defects with screened Coulombic tails that stretch to infinity*, causing the well-known divergence. In other words, the theory effectively treats a crystal containing "charged" defects as being truly charged, whereby this net charge would in principle vary with the charged-defect concentrations as functions of temperature. The consequences are exacerbated for charged defects at surfaces and 2D materials, where Coulombic potentials spill into the surrounding vacuum and extend to infinity. These scenarios are not in accord with physical reality because *neutral defects convert to charged defects by simply trading electrons with the energy bands*, i.e., they are merely *ionized* and *the crystal retains its charge neutrality*. The scenarios also violate a key principle of statistical mechanics according to which the Fermi level is determined by the equation representing charge neutrality (the charge of electrons in the conduction bands, holes in the valence bands, and all charged defects must add to zero). As a result, the conventional definition is applicable only when the Fermi level is controlled independently, e.g., by dopants.

The WZP paper recognizes that *for every charged defect there exists a neutralizing carrier in the energy bands, in accord with physical reality, which requires that every supercell is on average neutral*. The reformulated theory is based on a new definition, formally derived from rigorous statistical mechanics. The definition is flexible enough to handle undoped or doped materials or any other circumstance. There are no divergences and no long-range interactions that need to be dealt with, but the new definition is *not* simply another scheme to eliminate the divergence that is inherently present in the conventional definition of formation energies. The two definitions are fundamentally different: the conventional approach yields the formation energy of truly charged defects with Coulombic tails that stretch to infinity, whereas the WZP approach yields the formation energy of ionized defects in inherently neutral supercells. The WZP paper reported calculations for select defects in bulk materials. After publication of the WZP paper, two Comments were published.[43, 44] They did not address the relative validity of the two distinct theories. Instead, they argued that the differences in numerical results are small and focused on comparing implementation issues and numerical efficacy. Responses were also published,[45, 46] pointing out that the authors of the Comments did not dispute the fundamental issues that were raised by the WZP paper, which should be the appropriate criterion for judging the validity of two non-equivalent theories instead of computational issues. Both fundamental and computational issues, therefore, have remained open.

In this paper, we first give an extensive introduction on prior literature. We describe the evolution of the conventional approach and how the various complications arising from the divergences and the long-range Coulomb interactions, which are exacerbated for defects at surfaces and in 2D materials, have been treated, identifying open issues. In Section II, we describe the derivation of the formal definition of formation energies of neutral defects using statistical mechanics and how one obtains a practical prescription for practical calculations using either the Green's function approach or a supercell scheme. We then describe how the conventional theory of charged-defect formation energies is based only on an adaptation of the prescription for practical calculations of neutral-defect formation energies and describe in detail how the modified prescription fails to treat the physical electrons properly, which is the origin of the divergences and the other complications. We follow with a detailed construction of a theory of charged-defect formation energies in full accord with physical reality and the principles of the statistical mechanics of defects and electrons in semiconductors. We also derive the corresponding WZP prescription for practical calculations. Finally, we show that the conventional prescription for practical calculations can be derived from the WZP statistical-mechanics-backed prescription by invoking a series of approximations whose validity cannot be assessed *a priori*, i.e., uncontrolled approximations.

In Section III, we give calculation details that were adopted for the present paper. The objective of the present calculations is to compare WZP and jellium-based calculations of charged point defects at the same level accuracy and sophistications (choice of pseudopotential, exchange-correlation-functions, supercell sizes, k points, energy cutoffs, etc.). Since the jellium formation energy is in principle obtained by extrapolation of calculated values to infinite-supercell dimensions and model a posteriori "correction" schemes are only valid if the "corrected" formation energies converge to same value as the "uncorrected" formation energies, we use very large supercells and extrapolate the results of both methods to infinite supercells dimensions. The convergence rate of the WZP is excellent – the calculated values lie on totally flat straight lines in the case of charged defects in 2D materials, while the jellium values lie on curves that must be extrapolated to infinite-size supercells (no anomalous asymptotic behavior is found). The results show that formation energies of charged defects in 2D materials calculated by the jellium and WZP approaches differ by as much as ~1 eV, demonstrating that the two theories differ not only in the underlying fundamentals but also give very different numerical results.

In Section IV, we present extensive results on defects in diamond and 2D materials. The key results are: the numerical values of formation energies obtained by the two methods differ significantly, up to ~1 eV in monolayer h-BN. The net conclusion is that the two theories are not only different in that the WZP theory is in full accord with physical reality and statistical mechanics while the conventional approach is not, but the numerical results are also significantly different. In the absence of experimental standards of charged-defect formation energies, these results establish the WZP approach as the benchmark standard. Furthermore, we demonstrate that *the convergence behavior of WZP calculations is far superior to that of conventional calculations.*

Also in Section IV, we use the results of calculations to carry out a quantitative comparison of WZP and jellium features. Furthermore, we demonstrate that the vacuum regions in supercells for surfaces and 2D materials never cause any particular difficulties as the neutralizing charge, defined by band or dopant electrons, is naturally confined in the regions occupied by the material. Similarly, the electrostatic potential is naturally confined in the regions occupied by the material, with no unphysical potentials appearing in the vacuum regions, calling for *ad hoc* schemes to remove them, as is the case in the jellium approach. In particular, there are no restrictions as to how the vacuum region in the supercell should scale with the supercell dimensions.

Finally, in Section IV we also show explicitly that *the jellium scheme's removal of the divergence does not correspond to neutralizing the supercells by a uniform background charge density, which led to the moniker "jellium"*. Such a background density never enters the calculation anywhere [we show that a hypothetical uniform background charge density would not affect only the $V(\boldsymbol{G}=0)$ term of the defect potential, as has been claimed, but would alter the $V(\boldsymbol{G} \neq 0)$ terms as well, which does not happen in the DFT calculations]. This conclusion actually puts the "jellium" on a stronger footing, but the theory still treats truly charged defects with Coulombic tails that stretch to infinity. This scenario is not in accord with the physical reality of "charged" defects in crystals, which are merely ionized and require supercells that are explicitly neutralized by band carriers as in the WZP theory.

## II. Formation Energy Fundamentals

### II.A Neutral defects

The formation energy of a defect is defined within the context of statistical mechanics and thermodynamics. It is the average Gibbs-free-energy (GFE) cost per defect when a very large number of such defects is formed by thermal means in the dilute limit (negligible interactions between defects). Most defects of interest, e.g., vacancies, interstitials, and impurities require the removal of host atoms to a host-atom reservoir or the addition of host atoms or impurities from corresponding reservoirs. As a prototype example, let us consider vacancies in diamond. A statistically large number of atoms are removed, creating vacancies in the dilute limit at a certain energy cost per vacancy. The atom-removal process costs energy because bonds are broken. The atoms are placed "on the surface" for a gain of energy through new bond formation, but the details of the surfaces, e.g., whether they are flat or stepped, reconstructed or not, do not matter. We are dealing with very large numbers of atoms so that the net result is an extension of the bulk, i.e., the bulk crystal serves as the reservoir for the definition of the carbon chemical potential $\mu_C$ [Figures 1(a) and 1(b)]. Thus, the resulting GFE gain per vacancy is the GFE of an atom in diamond, calculated as $\mu_C = E_N^0/N$, where $E_N^0$ is the total GFE of a perfect-crystal unit cell containing $N$ atomic sites.

The formal definition of the neutral-vacancy formation energy can be expressed either in terms of a finite-size but macroscopic crystal containing $N$ atomic sites or in terms of an infinite crystal in which we define large periodic Born-von Karman (BvK) unit cells containing $N$ atomic sites, including $M_0$ vacant sites. The formation energy is then defined by

$$E_f^0 = \frac{\left(E_N^{M_0} + M_0 \mu_C\right) - E_N^0}{M_0}. \tag{1}$$

Here $E_N^{M_0}$ is the GFE of the finite crystal or a BvK cell containing $N$ atomic sites, $M_0$ of which are neutral vacancies distributed randomly in the dilute limit, and $E_N^0$ is the total GFE of the corresponding perfect crystal or BvK cell. This definition of formation energies underlies the fact that $E_f^0$ appears in the exponential $e^{-E_f^0/k_B T}$, where $k_B$ is the Boltzmann constant, which defines the concentration of the defects at temperature $T$ under equilibrium conditions.

From now on, we will focus on GFE's at $T = 0$, which correspond to enthalpies, namely the usual "total energies", computed using DFT. *Since we are in the dilute limit*, i.e., interactions between defects are negligible, we have two options for practical calculations. Both options can be captured by a single *prescription* deduced from Eq. (1):

$$E_f^0 = \lim_{N \to \infty} [(E_N^1 + \mu_C) - E_N^0] \tag{2}$$

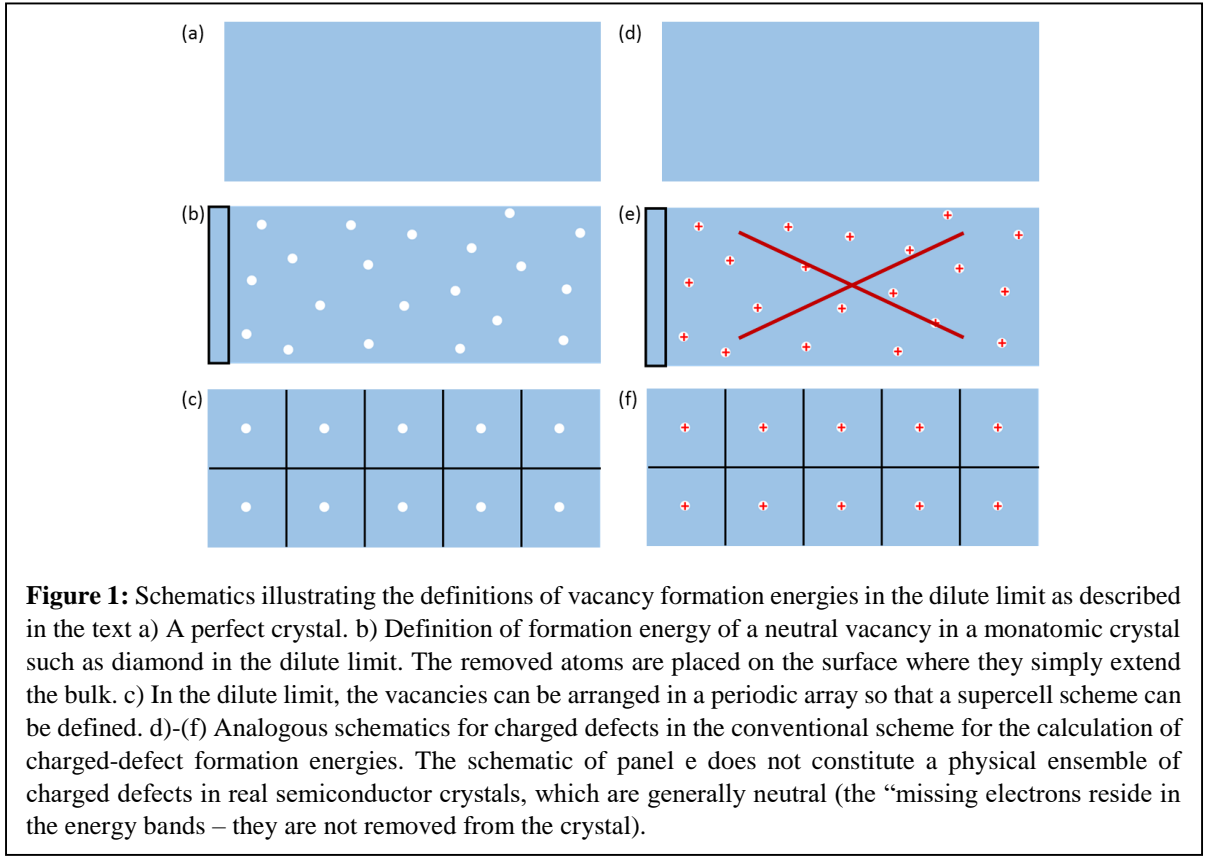

**Figure 1:** Schematics illustrating the definitions of vacancy formation energies in the dilute limit as described in the text a) A perfect crystal. b) Definition of formation energy of a neutral vacancy in a monatomic crystal such as diamond in the dilute limit. The removed atoms are placed on the surface where they simply extend the bulk. c) In the dilute limit, the vacancies can be arranged in a periodic array so that a supercell scheme can be defined. d)-(f) Analogous schematics for charged defects in the conventional scheme for the calculation of charged-defect formation energies. The schematic of panel e does not constitute a physical ensemble of charged defects in real semiconductor crystals, which are generally neutral (the "missing electrons reside in the energy bands – they are not removed from the crystal).

The case $N = \infty$ is precisely the prescription adopted in the original Green's-function calculations of formation energies in 1984[6-8] and amounts to a single vacancy in an otherwise perfect infinite crystal (calculations were done using a Gaussian basis set; "convergence" needs to be achieved with respect to the number of shells of Gaussians for which the defect potential has nonzero matrix elements).

Alternatively, the dilute limit allows us to order the $M_0$ neutral vacancies in $M_0$ periodically repeated "*supercells,*" each containing a single vacancy [Figure 1(c)]. Equation (2) then applies again, with $N$ now being the number of atomic sites in each supercell. In practice, one repeats the calculation using larger and larger supercells until the calculated formation energy "converges" to a stable value, signaling that interactions between defects in different supercells are negligible. If only very small supercells are practical and the calculated formation energy changes, it is difficult to determine a converged value by extrapolation because one does not know the curve's asymptotic behavior. It is then clear that a Green's-function and a converged supercell calculation, both of which are based on Eq. (2), would in principle give identical results for the formation energy of a neutral vacancy – a Green's-function calculation is in effect an infinite-supercell calculation. We emphasize that Eq. (2) is merely a *prescription* for computational purposes derived from the formal *definition*, Eq. (1).

The generalization of the above equations to compound materials and other defects and impurities is straightforward.[47, 48] One simply replaces $\mu_C$ with $\sum_{\alpha,\beta}(\mu_\alpha - \mu_\beta)$, where $\alpha$ runs over removed atomic species and $\beta$ runs over added atomic species. In particular, the supercell-based prescription is fully consistent with the statistical-mechanics-based definition of formation energies and has worked perfectly fine for the last 25 years. There are absolutely no issues with the theory for neutral defects.

**II.B Charged defects – The conventional approaches**

In contrast to neutral defects, *a statistical-mechanics-based derivation of the formation energy of charged defects has been lacking*. The early Green's-function calculations[6,7] for vacancies in Si simply adopted the following adaptation of Eq. (2), a practical prescription for calculations of neutral-vacancy formation energies:

$$E_f^q = \lim_{N\to\infty}[(E_N^{1q} + \mu_C + q\mu_e) - E_N^0], \quad (3)$$

with $N = \infty$. The "missing" or "excess" electrons simply contribute an energy $q\mu_e$, where $\mu_e$ is the electron chemical potential, namely the Fermi level $E_F$ relative to the valence-band edge (VBE). The calculated formation energies were displayed as functions of the Fermi level in a now-classic plot. No complications or ambiguities were encountered in this approach. Though the defect potential has a Coulomb tail that in principle extends to infinity, convergence of the formation energy of vacancies and self-interstitials in Si was achieved for a practical number of shells of Gaussian basis orbitals.[8]

The supercell approach effectively adopts Eq. (3), with $N$ being the number of atomic sites in each supercell. The generalization to arbitrary defects and materials again amounts to replacing $\mu_C$ by $\sum_{\alpha,\beta}(\mu_\alpha - \mu_\beta)$. However, though Eq. (3) and its generalization worked perfectly well in the 1980's in the case of the Green's-function approach, in which $N = \infty$, they are not suitable for calculations using the supercell approach. For finite-size supercells, the quantity $E_N^{1q}$ represents the total energy of an infinite periodic array of charged defects, as in Figure 1(f), whose overlapping Coulombic tails result in a divergence. More specifically, the $\mathbf{G} = 0$ component of the electrostatic potential, $V(\mathbf{G} = 0)$, is infinite.

Staying with our vacancies-in-diamond prototype, the state of the art for jellium calculations is as follows. The process of removing the divergence amounts to the following modification of Eq. (3):

$$E_f^q = \lim_{N\to\infty} \tilde{E}_f^q(N) = \lim_{N\to\infty}\left(\tilde{E}_N^{1q} - E_N^0 + \mu_C\right) + qE_F \quad (4)$$

where $\tilde{E}_N^{1q}$ is no longer a divergent quantity, subject to the strict condition that the modification vanishes at infinity. In the jellium scheme, it is recognized that the quantities $\tilde{E}_f^q(N)$ contains long-range interactions among defects, which are artifacts of finite-size supercells, whereby one must extrapolate to $N = \infty$, i.e., the $\tilde{E}_f^q(N)$ do not represent physical quantitites. Alternatively, *a posteriori* "corrections" remove the contributions of these long-range interactions, aiming for convergence as in the case of neutral defects, with even a single calculation using a relatively large supercell can be viewed as converged. Equation (4) is effectively modified to read

$$\tilde{E}_f^q = \lim_{N\to\infty}[\tilde{E}_f^q(N) + E_{corr}(N)] \quad (5)$$

where one must check the validity of the correction scheme by ensuring that the "uncorrected" $\tilde{E}_f^q(N)$ of Eq. (4) and the "corrected" $\tilde{E}_f^q(N) + E_{corr}(N)$ of Eq. (5) extrapolate to the same value, i.e., $\tilde{E}_f^q = E_f^q$. Though such tests have been carried for charged defects in bulk semiconductors, where it was found that the correction schemes work well in most but not all cases,[27] we are not aware of any such tests for charged defects at surfaces and in 2D materials, where the correction schemes become more complex and even show anomalous extrapolation behavior at very large supercell sizes.[37, 39] An alternative method for removing the divergence in the case of surfaces and 2D materials[13, 14] has apparently not attracted followers and we will not discuss it further.

The WZP critique of the conventional approaches is best illustrated using schematics. The schematics of Figured 1(d), 1(e), and 1(f), which completely parallel the schematics of Figures 1(a), 1(b), and (c) for neutral vacancies show that the atom removal is handled correctly, as in the case of neutral vacancies. The problems with Eq. (3) arise with the treatment of electrons. Though Eq. (3) handles the energy cost of converting neutral defects to charged defects by recognizing that, on average, the electrons are traded with a hypothetical energy level at the Fermi level, the physical electrons are not treated correctly: Equation (3) assumes that an ensemble of neutral vacancies in the dilute limit in either an infinite crystal or a large BvK cell are converted to positively charged vacancies by removing $q$ electrons per vacancy from the crystal altogether, leaving the crystal truly charged as in the schematic of Fig. 1(e). The concentration of charge corresponds to the concentration of charged defects, which of course is a function of temperature. It is the schematic of Figure 1(e) that is adapted to Figure 1(f) to arrive at the standard supercell configuration that gives rise to the divergence. In the infinite-supercell limit, we recover the Green's function scenario of a single charged defect in an otherwise infinite perfect crystal.

The large red X we placed on Figure 1(e) is meant to convey the fact that the schematic contradicts physical reality: neutral defects are converted to charged defects by trading electrons with the energy bands, always leaving the crystal neutral, in accord with the statistical mechanics of electrons in semiconductors, whose principal tenet is that the Fermi level is determined by imposing charge neutrality. Indeed, Eq. (3) and its generalization to defects in compound semiconductors can only be used when the Fermi level is controlled independently by some means, e.g., by dopants, whose effect on the formation energy of the defect of interest can be ignored. If some native defects, e.g., vacancies, is the sole defect under consideration, as is often the case in undoped materials, Eqs. (3)-(5) would stand alone with no way to determine the Fermi level.

The net conclusion is that the schematic of Figure 1(e) does not correspond to a physical ensemble and one cannot derive an analog of Eq. (1) for charged defects in this fashion. The above observations regarding the basis of the conventional definition of charged-defect formation energies form the motivation for the WZP reformulation of the problem of charged-defect formation energies, in a way that is fully consistent with the statistical mechanics of both atoms and electrons in semiconductors. In Section II.C, we give an expanded derivation of this reformulation. We emphasize that we will present a fundamental reformulation of the theory of charged-defect formation energies and derive a formal definition using statistical mechanics and a corresponding prescription for practical calculations in both the Green's-function and supercell schemes. The formal definition and corresponding prescription by their very nature do not entail a divergence, but the new formulation is not in any way yet another way of eliminating the divergence in Eq. (3), *as Eq. (3) is discarded altogether and never invoked*.

After presenting the WZP, statistical-mechanics-based reformulation of the theory of charged defects in Section II.C, we will show in Section II.D that the conventional *prescriptions*, Eqs. (3)-(5), can actually be derived from the new *formal definition* only if one invokes specific approximations whose consequences cannot be assessed a priori, i.e., uncontrolled mathematical approximations. In Section IV, we will present extensive new calculations of charged-defect formation energies and demonstrate that, though the differences between the jellium and WZP approaches tend to be relatively small in bulk materials, the differences in the case of defects in 2D materials are very substantial in both numerical values and in the way the calculated formation energies vary as a function of supercell size (unlike the conventional values, the WZP-derived values are physically meaningful at any supercell size). After the presentation of numerical results, we will revisit one more time key differences between the two approaches and demonstrate their numerical significance. Our net conclusion in the end is that the new theory is the *de facto* standard for accurate and reliable calculations.

**II.C Statistical mechanics of charged defects – The WZP formulation**

We go back to the derivation that we gave for the definition of the formation energy of *neutral defects* using vacancies in diamond as a simple prototype. In that case, we dealt with the statistical mechanics of removing atoms from a crystal and, more generally, adding or replacing atoms with impurities. In order to handle formation energies of charged defects we must engage with the statistical mechanics of electrons and holes in the presence of defects and possibly dopants.[42] In all cases, at a finite temperature $T$, the crystal is overall neutral and carries a concentration $n$ of electrons in the conduction bands, a concentration $p$ of holes in the valence bands, concentrations $C^q$ of the different charge states $q$ of a defect ($q = 0, \pm 1, \pm 2, \ldots$), and concentrations $C_{dop}^{q'}$ of ionized dopants carrying charge $q'$. The Fermi level is then determined by charge neutrality,

$$p - n + \sum_q q C^q + q' C_{dop}^{q'} = 0. \qquad (6)$$

The distribution of the electrons and holes on the energy axis at temperature $T$ is controlled by the F-D distribution function. In particular, we have

$$n = \int_{\epsilon_{CBM}}^{\infty} \frac{g(\epsilon) d\epsilon}{e^{(\epsilon - E_F)/k_B T} + 1} \qquad (7)$$

and

$$p = \int_{-\infty}^{\epsilon_{VBM}} \frac{g(\epsilon) d\epsilon}{e^{(E_F - \epsilon)/k_B T} + 1} \qquad (8)$$

where $g(\epsilon)$ is the density of one-electron states, which for our purposes can be approximated by effective-mass expressions.

At this point, we need to distinguish between the case of defects in an undoped versus doped semiconductor. We first describe the case of undoped diamond, for which the derivations are simpler.

*i. Charged defects in undoped semiconductors*

Let us again consider vacancies in diamond for concreteness. We start with the schematic of Figure 1(b), where a total of $M_0$ *neutral* atoms are missing from a large Born-von-Karman-type unit cell and are accommodated in an extension of the bulk, which defines the carbon atom chemical potential. At any finite temperature, a fraction of the vacancies are *positively "charged", which means that one or more bound electrons transfer to the conduction bands,* and a fraction of the vacancies are *negatively "charged", which means they have captured one or more electrons from the valence bands, leaving behind holes*. <u>The crystal is always neutral overall as no electrons are physically removed from it.</u> This is the key observation that leads to an alternative definition of charged-defect formation energies.[42]

To derive the formation energy of charged vacancies in the $q$ charge state, we simply start with all neutral vacancies and convert $M_q$ of them with concentration $C^q$ into the $q$ charge state. The key point here is that, according to statistical mechanics, *electrons are effectively traded between the Fermi level as a hypothetical energy level* serving as the chemical potential, *even though physically they are traded with the energy bands*. Statistical mechanics mandates that both these principles are incorporated in the definition of formation energies. The conventional prescription for supercell calculations of charged-defect formation energies, Eqs. (3)-(5), which are simply adaptations of Eq. (2), the prescription for supercells calculations of neutral-defect formation energies derived from a statistical-mechanics-based definition, honor only the first of these two principles, but fail to treat the physical electrons in an appropriate manner. By invoking both principles, the formation energy $E_f^q$ for $q > 0$ is defined by

$$E_f^q = E_f^0 + \frac{E_N^{M_q(q)} - E_N^{M_q(0)}}{M_q} - \frac{1}{M_q}\left\{\int_{\epsilon_{CBM}}^{\infty} f_{FD}(E)(\epsilon - E_F)g(\epsilon)d\epsilon - \int_{\epsilon_{CBM}}^{\infty} f_{FD}^0(E)(\epsilon - E_F^0)g^0(\epsilon)d\epsilon\right\} + qE_F. \quad (9)$$

Here $E_N^{M_q(q)}$ is the free energy of our BvK-like large cell with $N$ atomic sites, $M_q$ positively charged vacancies (the parentheses around $q$ signify that the $|q|$ electrons per defect are present in the energy bands of the BvK cell), and $qM_q$ extra electrons in the conduction bands distributed together with the intrinsic thermal electrons as per the F-D distribution function, while $E_N^{M_q(0)}$ is the total free energy of the same large cell containing $M_q$ neutral vacancies. The remaining terms in Eq. (9) represent the fact that, though the $qM_q$ electrons that left the defects are physically in the conduction bands and participate in determining $E_N^{M_q(q)}$, their statistically effective contribution to the energy is *as if* they occupy a hypothetical state at the Fermi level. The key point for this dichotomy in treating the ionized electrons is that their physical presence in the conduction bands has a real effect on lattice relaxation and *all the other electrons* that contribute to $E_N^{M_q(q)}$ [the first integral in Eq. (9) is the energy of all the electrons in the conduction bands when the charged defects are present with the Fermi level at $E_F$; the second integral is the energy of the thermal electrons in a perfect BvK cell, in which the Fermi level is $E_F^0$].

In order to derive a prescription for practical calculations of charged-defect formation energies, we proceed as follows. In the general ensemble of vacancies depicted in the schematic of Figure 1(b), we can assign a Voronoi polyhedron to each vacancy. Since the crystal is generally neutral, though fluctuations occur, on average, over macroscopic time scales, *every Voronoi polyhedron can be viewed as neutral* containing either a neutral vacancy or a charged vacancy with charge $q$ plus the corresponding $|q|$ band carriers (the persistent neutrality of atoms, even in so-called ionic crystals and other cases where ions are expected to exist has been documented extensively[49-52]. No electrons are ever missing or added to the crystal. In the dilute limit, instead of drawing Voronoi polyhedrons around the vacancies, we can order the vacancies in sufficiently large periodic supercells, i.e., we go from the schematic of Figure1(b) to that of Figure 1(c), but we now view each supercell as containing either a neutral vacancy or a "charged" vacancy, with the corresponding neutralizing carriers being present in the energy bands (the physical reality, which is enshrined in the statistical mechanics of defects, is that the vacancies are constantly altering their charge state while the average concentration of each charge state remains relatively constant; it is this average concentration that is determined by the free energy of formation of each charge state).

Similar considerations apply for $q < 0$. The reduction of Eq. (9) and the corresponding equation for negatively charged defects to the analog of Eq. (2), i.e., a *prescription* for practical calculations, is the same for both positive and negative $q$:

$$E_f^q = \lim_{N \to \infty} [((E_N^{1(q)} - q\epsilon_b + qE_F) + \mu_C) - E_N^0], \qquad (10)$$

where we used Eq. (1) to eliminate $E_f^0$. The parentheses around $q$ in $E_N^{1(q)}$ indicate that the $q$ electrons removed from each vacancy to make it positively charged are placed at a representative state at energy $\epsilon_b$ in the conduction bands. Similarly, the $|q|$ electrons added to each vacancy to make it negatively charged left holes at a representative state at energy $\epsilon_b$ in the valence bands [we are distinguishing from the notation $E_N^{1q}$ in Eq. (3) where vacancies in the $q$ charge state are not compensated by carriers in the energy bands, resulting in a supercharged crystal]. Note that we neglect thermal electrons and holes, but in principle their role can be estimated by introducing an extra electron and an extra hole in supercells whose size would best match the concentration of thermal electrons and holes at a given temperature.

As is the case with Eq. (2), Eq. (10) applies to both Green's-function and supercell-type calculations as follows. For the Green's-function calculation, $N = \infty$, and one deals with a single vacancy either with $q$ of its own electrons occupying a representative state at $\epsilon_b$ in the conduction bands or with $|q|$ extra electrons added to vacancy bound states, leaving $|q|$ empty states at a representative energy $\epsilon_b$ in the valence bands. Similarly, in a supercell formulation, we order the vacancies, one in each periodic supercell, but there are $q$ electrons for each positively charged vacancy occupying a representative state at $\epsilon_b$ in the conduction bands, i.e., each supercell is always neutral. For negatively charged vacancies, there are $|q|$ empty states at a representative energy $\epsilon_b$ in the valence bands, resulting in neutral supercells. Calculations are to be done for supercells containing $N$ atomic sites, with $N$ made successively larger, seeking convergence to a stable value. In this fashion the supercells are always neutral and the infinite-supercell-limit formation energy would automatically be equal to the corresponding Green's-function value. Note that we neglect thermal electrons and holes, but in principle their role can be estimated by introducing an extra electron and an extra hole in supercells whose size would best match the concentration of thermal electrons and holes at a given temperature through F-D statistics.

At typical temperatures, there are no hot electrons or hot holes, so that in Eq. (9) the "representative" state at $\epsilon_c$ in the conduction bands can be taken to be the lowest energy state in the conduction bands of the supercell containing the vacancy, which may evolve with supercell size. Alternatively, one can use higher-energy states, sampling the integral in Eq. (9). If the resulting formation-energy differences are not negligible, one should take an average over several choices of $\epsilon_c$ [in principle, one should use large supercells with $M_q$ ionized vacancies and do the integral of Eq. (9), then divide the result by $M_q$]. The Fermi energy and formation energies are ultimately determined by a simultaneous solution of Eqs. (6)-(8) and (10). The generalization of Eq. (10) to arbitrary defects in all materials is straightforward, as in the conventional theory, i.e., the term $\mu_{Si}$ is replaced by $\sum_{\alpha,\beta}(\mu_\alpha - \mu_\beta)$.

*ii. Charged defects in doped semiconductors*

In a doped semiconductor where the Fermi level is controlled by the dopant concentration, the formation energy of charged defects depends on whether the material is *n*-type or *p*-type for the following reason. Positively charged defects in *n*-type material are treated as above, i.e., the ionized electrons are placed in the conduction bands. For negatively charged defects in *n*-type material, however, the extra electron(s) do not originate in the valence bands leaving a hole behind. Instead, they are more appropriately viewed as donor electrons. Equation (10) still applies, but, for $q = -1$, $E_N^0$ should be computed in large supercells containing one dopant impurity with its electron in the conduction bands, while $E_N^{1(q)}$ is computed in a supercell containing the defect plus a dopant impurity, with the dopant electron transferred to a defect bound state (since dopant levels are typically shallow, effective-mas like, one can simply place the electron for the $E_N^0$ calculation either in the lowest-energy or second-lowest-energy empty state and the difference should be negligible). A similar procedure needs to be used for positively charged defects in *p*-type material, where the ionized electron is not placed in the conduction bands but in an acceptor level.

*iii. Key features of the new prescriptions for practical calculations of charged-defect formation energies*

The above prescriptions for supercell calculations, derived from the formal statistical-mechanics-backed definition of charged-defect formation energies, are an internally consistent set that is directly applicable to defects in bulk solids, at surfaces, and 2D semiconductors, with or without dopants controlling the Fermi energy. No worrisome or uncontrolled approximations have been made as no divergences ever appear anywhere. The vacuum regions in supercells for surfaces and 2D materials never cause any particular difficulties as the neutralizing charge, defined by band or dopant electrons, is naturally confined in the regions occupied by the material. Similarly, the electrostatic potential is naturally confined in the regions occupied by the material, with no unphysical potentials appearing in the vacuum regions, calling for *ad hoc* schemes to remove them. In particular, there are no restrictions as to how the vacuum region in the supercell should scale with the supercell dimensions.

One additional point is worth noting. The calculated formation energy at any size supercell can be viewed as the true formation energy at the corresponding defect concentration (one defect per supercell volume, which can easily be translated to physical units) in the limit when pairing or clustering is negligible. Thus, in the absence of convergence, one does not necessarily have to attempt an extrapolation to infinity, which, as pointed out in Section II.B, may be fraught with uncertainties.

**II.D Relation of the jellium scheme to the WZP statistical-mechanics-backed definition**

We are now in position to examine how the conventional supercell scheme for charged-defect formation energies [Eqs. (3)-(5)], especially the widely used jellium scheme, relates to the formal definition that we just derived using statistical mechanics. We already pointed out in Section II.B that Eqs. (3)-(5) of the jellium scheme handle the atom-removal or more generally the atom-replacement part the same way as for neutral defects, which is in accord with statistical mechanics as described in Section II.A. The conversion of neutral defects into charged defects, however, amounts to an *unphysical* process, as electrons are removed from the crystal or added to the crystal, leaving the crystal supercharged as shown schematically in Figures 1(e) and 1(f). In Section II.B, we identified the principles of statistical mechanics that are violated when electrons are removed from the crystal or added to the crystal in order to form charged defects. Our task here is to show that Eqs. (3)-(4), which are the cornerstones of the jellium scheme, can be derived from the statistical-mechanics-backed WZP definition of charged-defect formation energies by imposing certain approximations and then assess their implications.

The key "approximation" that must be imposed to go from the rigorously derived Eq. (10) to Eq. (3), which serves as the cornerstone of the jellium scheme, is to drop the neutralizing band carrier(s) in the definition of $E_N^{1(q)}$ and the term $-q\epsilon_b$ in Eq. (10). These two contributions do not cancel. After all, $E_N^{1q}$ is divergent. Moreover, the neutralizing carriers enter the calculation of $E_N^{1q}$ in several distinct ways. During the self-consistency loop for the electronic structure calculation in the defected supercells, the WZP electron density $\rho(\bm{r})$ for our prototype positively charged vacancy in diamond is given by

$$\rho(\bm{r}) = \rho^+(\bm{r}) - e|\psi_b(\bm{r})|^2. \tag{11}$$

The second term in this expression affects all the other electrons and hence the total electron density, which determines $E_N^{1(q)}$. Furthermore, the band wave function $\psi_b(\bm{r})$ may exhibit a degree of localization around the ionized defect while still being Bloch functions further out. We recall that the continuum solutions of a hydrogen atom *in vacuum* are not plane waves, but Whittaker functions,[53, 54] which have a degree of localization around the proton and approach plane waves at some distance from the proton. Clearly, any degree of localization of the neutralizing band carriers described by $\psi_c(\bm{r})$ in Eq. (11) would impact the full $\rho(\bm{r})$, the resulting total energy, and concomitant lattice relaxations in ways that cannot be easily assessed without performing a full WZP calculation. Since dropping the neutralizing band electrons leads to Eq. (3) where $E_N^{1q}$ is divergent, we conclude that such an "approximation" is uncontrolled. Furthermore, one intentionally ends up calculating the formation energy of truly charged defects as in the schematics of Fig. 1(e) and 1(f), representing a crystal carrying a significant concentration of charge that varies with temperature, which is not in accord with physical reality.

In addition to dropping the band electrons, the jellium scheme entails additional approximations in the process of removing the divergence in the resulting defect potential. As we discussed in the Introduction, the jellium scheme asserts that adopting the condition $V(\bm{G}=0)=0$ is equivalent to introducing an "inert" homogeneous background charge density that effectively neutralizes the supercells. To scrutinize the nature of this scheme, we note that we can go from Eq. (10) to Eq. (4) by the following procedure. In the self-consistency loop for the charged-defect electronic-structure calculation, the contribution to the total charge density $\rho(\bm{r})$ from the neutralizing band electron is first replaced by a uniform density ("a neutralizing background charge density") as if the band electron occupies a plane-wave state normalized in a single supercell:

$$\rho(\bm{r}) = \rho^+(\bm{r}) - e|\psi_c(\bm{r})|^2 \to \rho^+(\bm{r}) - \frac{e}{\Omega}. \qquad (12)$$

At this point, just as in the WZP scheme, there is no divergence and one is allowed to set $V(\bm{G}=0)=0$. It would then appear that $e/\Omega$ goes to zero in the infinite-supercell limit, whereby one can drop it in finite-supercell calculations and then extrapolate the results to the infinite-supercell limit. The jellium scheme, which seeks to determine the formation energy of truly *charged* defects as in the schematic of Figure 1(f) does that and therefore appears to be fully justified, as long as the finite-supercell results are not viewed as physically meaningful. One can even argue that, in large supercells, one expects the effect of $e/\Omega$ to be relatively small on the total $\rho(\bm{r})$, whereby jellium calculations should be fine. We already addressed this question: The contribution of the band electron to $\rho(\bm{r})$, namely $e|\psi_c(\bm{r})|^2$, and its effect on the local lattice relaxation and total energy cannot be assumes *a priori* as negligible. The "approximation" would be uncontrolled. The new calculations presented in Section IV below demonstrate that the effect of band electrons on the numerical differences between WZP and jellium calculations at the same level of approximation can be quite large.

There are additional issues, however, with the process of going through Eq. (12), dropping the term $e/\Omega$, which generates a divergence in the potential, and then setting $V(\bm{G}=0)=0$.

1. One is not at liberty to simply invoke the last form of $\rho(\bm{r})$ in Eq. (12) because the rules of self-consistency require that the $\rho(\bm{r})$ must be constructed out of the wave functions of the occupied states. There really exists no rationale for invoking such a step.

2. Even if one accepts Eq. (12), the practice of dropping the term $e/\Omega$ in order to describe a truly "charged", as opposed to an ionized defect, while retaining its presumed effect on the potential, namely setting $V(\bm{G}=0)=0$, is tantamount to violating Poisson's equation at every step of the self-consistency loop: the charge density is $\rho^+(\bm{r})$ in a charged supercell, while the non-divergent potential is that of a neutralized supercell [a uniform background charge density $e/\Omega$ of the correct sign has a divergence that cancels exactly the divergence arising from $\rho^+(\bm{r})$]. The net result is that the DFT expressions for the effective Hamiltonian and total energy, instead of being functionals of a single electron density, they are actually evaluated using a mix of $\rho^+(r)$ and the form of $\rho(r)$ to the right of the arrow in Eq. (12). More specifically, the Hartree energy is expressed in terms of the product $\rho^+\rho$; the mix carries into the kinetic energy, whereas the exchange-correlation energy is expressed in terms of $\rho^+$. One must then assert that formation energies calculated in this fashion using finite supercells can be relied upon to produce a correct result if extrapolated to the infinite-supercell limit. We note that one can equally well perform finite-supercell calculations using the potential derived from the last form of $\rho(r)$ instead of just using the condition $V(G = 0) = 0$. The resulting formation energies, which are likely to be different, should extrapolate to the same value because $e/\Omega \to 0$ in the infinite-supercell limit. To our knowledge, no one has reported such calculations. We also note that *a posteriori* corrected values do not represent physical formation energies either if they vary with supercell size, and are judged valid only if they produce the same infinite-supercell limit[27]. Their only role is to speed up the process of obtaining a final result. In contrast to all the above, in the WZP formulation, *all* terms in the total energy are evaluated as functionals of $\rho(r)$ of Eq. (11) and Poisson's equation and the rules of self-consistency are never violated. Formation energies calculated by the WZP method using finite supercells are physically meaningful for the corresponding concentration $1/\Omega$ in the absence of defect pairing and clustering.

3. Finally, even if we accept all of the jellium "approximations" and the extrapolation to the infinite-supercell limit as a reasonable procedure, its target is to produce a formation energy for a charged defect in the sense that one or more of its electrons are removed entirely from the crystal or one or more electrons are added to the crystal, which does not correspond to physical reality. The derivations in Section II.C leave no doubt that electrons are not removed from or added to the crystal to produce "charged" defects. Instead, they are traded with the energy bands.

It is worth noting that, though for charged defects in bulk crystals it is difficult to assess the impact of the jellium procedures, which are employed in order to calculate formation energies of truly "charged", as opposed to merely ionized defects, the case of charged defects at surfaces or in 2D materials is easier to analyze: the "approximation" of Eq. (12) is clearly unacceptable at the outset because $\psi_c(r)$ *is naturally confined within the material*, whereas the uniform density $e/\Omega$ *is uniformly distributed in both the material and the vacuum region in each supercell*. The mixed use of $\rho^+$, which is confined in the region of the material, and $\rho$, which has a uniform component in the vacuum regions, may cause irreparable harm to the "uncorrected" formation-energy values, which are presumed to extrapolate to a correct infinite-supercell limit. Moll et al.[14] have pointed out that the hypothetical neutralizing background charge density is responsible for the divergence that occurs if the vacuum region is increased while keeping the in-plane supercell dimension constant. Scaling the supercell in all three dimensions removes the divergence, but the effect of mixing $\rho^+$ and $\rho$ on the "uncorrected" formation-energy values remains uncontrolled. One must then rely entirely on the *a posteriori* finite-supercell corrections to produce a correct extrapolation. As pointed out by Noh et al. and Komsa et al. the inhomogeneous dielectric screening that is inherent when vacuum regions are present in the supercells, the *a posteriori* corrections exhibit an anomalous behavior at large supercells, requiring a high-order polynomial to fit the calculated values. These authors assumed that the DFT-calculated formation energies obey the same polynomial at large supercells, but this assumption has not been tested by calculating jellium formation energies using very large supercells. In Section IV.B we will present such results for charged defects in monolayer h-BN and monolayer MoS$_2$ using both jellium and the WZP method. The WZP results converge quickly while the jellium results do not show any anomalous behavior that requires a high-order polynomial for extrapolations. Thus, the ability of the jellium method to produce satisfactory numerical results by using relatively small supercells has not yet been demonstrated when vacuum regions are present in the supercells.

In Section IV, we present new calculations for defects in both bulk and 2D materials and show that the differences between the two approaches can indeed be relatively small for defects in bulk materials, but they are very large in monolayer h-BN, in both numerical values and convergence rates. In addition, in Section IV.C, we will use numerical results for particular defects in both bulk and 2D materials to discuss further the issues we raised in this section on the validity of the jellium scheme. In the aftermath of presenting numerical results we will re-engage in a comparison of the numerical features of the jellium and WZP method. At that point we will demonstrate that, despite the unsubstantiated claims to the contrary, the jellium method's setting $V(\boldsymbol{G} = 0) = 0$ *does not correspond to introducing the effect of a "neutralizing homogeneous background charge density on the supercell potential"* as the moniker "jellium" indicates. A strong indication that the superclass are not neutralized when $V(\boldsymbol{G} = 0)$ is set to zero is the fact that the Coulombic tails persist unmodified as their long-range interactions are removed by a model set of Coulomb interactions for which the $V(\boldsymbol{G} = 0)$ term is also set to zero. The realization that the jellium scheme entails no neutralizing background charge density actually sets the scheme on a firmer mathematical footing (Poisson's equation is not violated), but the method still calculates the formation energy of charged defects in the absence of neutralizing band carriers, which violates the principles of the statistical mechanics of electrons in semiconductors and is contrary to physical reality.

### III. Calculation details

In this section, we report results of calculations performed at the same level of accuracy (same pseudopotentials and exchange-correlation functional, same single k point for the Brillouin -zone sums, and same energy cutoff in the plane-wave basis set, using the WZP and jellium approaches. For each defect, we employ several fairly large supercells so a straightforward extrapolation to infinite-dimension supercell is possible using fits of the calculated values to low-order polynomials. We pursue such calculations because the standard of reference for the jellium method is the extrapolated value (in the WZP calculations for charged defects in 2D materials, the calculated formation energies actually converge so that no extrapolation is necessary). We do not engage in *a posteriori* corrections for jellium finite-supercells results because the corrections are valid only if the "corrected" formation energies extrapolate to the same value as the "uncorrected" formation energies. In this fashion, we get a direct and unambiguous comparison of the numerical results obtained by the two approaches for each-charged defect formation energy. Our sole purpose is to examine how different are the numerical results obtained by the two methods.

We have performed calculations of charged-defect formation energies in both 3D and 2D materials. In the case of 3D materials, we present the case of vacancies in diamond with $q = -2$ to show that indeed, in some cases, conventional jellium calculations give results that can be quite close to those of the WZP approach. We shed light on the origins of such results by examining the localization of the band electrons. We then report results for several defects in monolayer h-BN where we show that the conventional jellium formation-energy values can be very different from those obtained using the WZP method. We again analyze the origin of the differences. In this Section we describe the calculation details and in the next Section we report the results.

Density functional theory, as implemented in the computer code VASP[55, 56] (version 5.4.1), is employed for defect calculations. For defects in diamond, we use cubic supercell structural models with lattice sizes between 14 and 28 Å (up to 4096 atoms). For defects in monolayer BN, we use hexagonal supercells with lateral sizes between 20 and 80 Å (up to 2048 atoms). *The calculated formation energies in BN actually converge very fast* so that the larger supercells are only needed for the jellium calculations, for which the calculated formation-energy values must be extrapolated to infinite-supercell dimensions. We performed jellium calculations for very large supercells to in order to check the assumption that the "uncorrected" formation energies obtained by using very large supercells exhibit a 5$^{th}$-order polynomial or other more complex behavior, matching the behavior of the correction models. We found no such behavior. For one of the defects we display fits to the calculated values using high-order polynomials to show that they are not appropriate.

In all calculations, we treat the outer valence electrons of each atom explicitly and use the projector augmented wave (PAW) method to capture the effect of core electrons.[57, 58] A plane-wave basis with a well-converged energy cutoff is used. The cut-off energy is 400 eV for both diamond and monolayer h-BN. For exchange-correlation, we employ the local density approximation (LDA)[59, 60] for diamond calculations in order to directly compare to previous work; however, we use the Perdew-Burke-Ernzerhof (PBE) generalized gradient approximation for h-BN. Calculations using hybrid, meta-GGA, or hybrid meta-GGA exchange-correlation functionals would give more accurate values because of LDA and GGA band gaps are typically much smaller than experimental values, but, the new-generation functionals are computationally very taxing. *Our focus here is not to report the most accurate formation energies in any one material, but to calculate formation energies at the same level of accuracy using the WZP and jellium approaches.* Since the two approaches are not in any way equivalent and the jellium approach is derivable from the formal statistical-mechanics-backed WZP definition by invoking approximations whose accuracy cannot be assessed a priori, substantial differences between the calculated values establish the WZP approach as unambiguously the standard and the method of choice for future work.

For integrations over the Brillouin zone, it has long been understood that one can accelerate convergence by using special k-points that avoid Brillouin zone edges.[61] Indeed, we find one special k-point is well-converged for the large supercells considered here. Finally, spin polarization is used as needed and alters the charged defect formation energies by less than 0.1 eV.

We performed charged-defect calculations in parallel using both the jellium and WZP methods. In the jellium approach the lowest energy eigenvalues are occupied automatically, leaving the supercells charged by the net charge $q$ of the defect being examined. A Gaussian smearing, with a 0.05 eV spread, is used to determine occupations. The divergence of the Coulombic interaction term is removed by setting the average electrostatic potential equal to zero [$V(\boldsymbol{G}=0)=0$]. In the WZP method, on the other hand, the supercells ae automatically charge-neutral. Electrons are moved from defect states to band states and vice versa by fixing the eigenvalue occupations (constrained LDA or GGA) using the VASP NBANDS feature. The band occupations are fixed at zero or one throughout the entire electronic self-consistent calculations. While the occupations are fixed, the charge density and wave functions are allowed to relax until a self-consistent total charge density is reached. The considered defects have defect levels that lie deep within the theoretical band gap. While the defect-level occupancies are fixed and correspond to the nominal charge q of the defect, the computer code relaxes all electronic states allowing for mixing of the neutralization band states and the defect states, as would be the case in a physical system. Calculations are typically done by placing the band electrons at the lowest conduction level or the highest valence level in the supercell, as long as these levels are not true localized defects states. We also report calculation where the band electron in the conduction bands is placed at a series of levels above the conduction-band minimum.

For a defect in a charge state $q$, the formation energy in a crystalline system, $E_f^q$ is determined by a set of total-energy supercell calculations. In the jellium approach, when exchanges of atoms with atomic reservoirs are involved, the supercell version of Eq. (4) is given by

$$E_f^q = E_1^q - E_0 + \sum \mu_i \Delta n_i + q\, E_F, \qquad (13)$$

where $\Delta n_i$ is the change in the number of atoms of type $i$ and $\mu_i$ is the chemical potential for atom type $i$. In Eq. (13) we did not include finite-supercell corrections whose purpose is to expedite convergence because, as discussed above, it is more appropriate for the purposes of this paper to compare values obtained by the two methods at the same level of accuracy.[27, 62 26, 58] In the WZP method, when exchanges of atoms with atomic reservoirs are involved, the supercell version of Eq. (10) is generalized to[42]

$$E_f^q(E_F) = E_1^q + q(E_F - E_b) + \sum \mu_i \Delta n_i - E_0. \qquad (14)$$

As has been the general practice, we plot the calculated formation energies of both the WZP and jellium aproaches as functions of $L^{-1}$, where $L$ is the characteristic length of the periodic supercell. Therefore, when convergence is not occurring, the y-intercept of the extrapolation of $E_f^q$ vs $L^{-1}$ represents the dilute-limit formation energy, which we refer to as $E_f^\infty$. We generally use either linear or quadratic extrapolations as appropriate. In the WZP method, extrapolations are needed only in the case of defects in bulk materials, whereas the "uncorrected" jellium results require extrapolations in all cases. We will discuss the issue of higher-order-polynomial fits of calculated values, as it was discussed in the introduction in the context of papers by Noh et al[37] and Komsa et al.,[39] using our calculated results for very large supercells for defects in BN.

The present calculations, based on LDA or GGA, suffer from the fact that band gaps are underestimated, but the results are adequate to prove that, at the same level of approximation (functional choice, k-points for Brillouin zone integration, etc.), the two non-equivalent definitions of charged-defect formation energies yield substantially different values for defects in h-BN. As computers become faster, calculations using hybrid functionals should become possible for higher accuracy.

One additional point remains to be addressed. In jellium calculations, the position of the Fermi level in Eq. (13) is traditionally taken to be relative to the valence-band maximum (VBM) obtained from the defect-free unit cell calculation. To facilitate comparison between jellium and WZP calculations, for jellium calculations we use the VBM for negatively charged defects and the conduction-band minimum (CBM) for positively charged defects. The reason for these choices is the presence of the term $q(E_F - E_b)$ in Eq. (14), the WZP expression for formation energies. It is then most convenient to use $E_F = E_{CBM}$ for positively charged defects and $E_F = E_{VBM}$ for negatively charged defects. In this fashion, the value of the band gap, which can change with supercell size, never enters the calculation of formation energies. It should be pointed out, however, that this practice is necessary for the purposes of this paper whose focus is a comparison of the WZP and jellium approaches at a fundamental level. When one seeks to calculate the most accurate formation energies of specific defects, we remind the reader that the WZP calculation of charged-defect formation energies depends on whether the crystal is undoped, doped n-type, or doped p-type; in the undoped case, the Fermi level must be determined self-consistently with the formation energies of all charge states of the defect (see discussion in Section II.C). We defer such calculations to future papers.

Finally, we note that for the calculations for defects in h-BN, none of the supercells used are a 3N multiple of the primitive unit cell. Due to the use of a single k-point and band folding, the 3N×3N supercell formation energies lie slightly off the curve defined by other supercells sizes for both the jellium and WZP results. These issues would be absent if the number of k points is converged in all cases.

In order to analyze the differences in the jellium and WZP formation energies, we find it useful to examine the electron density of pertinent eigenstates, especially the band state occupied by the neutralizing carrier in the WZP scheme. The electron density of a particular state is calculated using the LPARD flag in VASP. One direction is chosen for the plotting variable and the density is averaged over the other two perpendicular directions. The resulting "plane-averaged density" is oscillatory due to the effect of the crystal structure. By averaging over the crystalline lattice planes, we can quantify the long-range variation of the state by removing the short-range oscillations. The long-range variation in the "lattice-average density" can be further quantified by the min/max ratio $R_{mm}$, i.e. the ratio of the minimum to the maximum average density values.

## IV. Results of calculations

### IV.A Charged defects in bulk semiconductors

The unrelaxed vacancy in diamond is a prototypical defect used in numerous studies to test methods for accelerating the convergence of jellium calculations.[21, 27] Diamond is a good test system for charged-defect calculations because of its low dielectric constant, which results in slow convergence of the jellium calculations. In 2012, Komsa, Rantala and Pasquarello (KRP) published LDA formation energies for charged vacancies in diamond using bulk supercells with 64, 216 and 512 atoms.[27] We performed such calculations for vacancies in diamond in both the jellium and WZP schemes using supercells with 512, 1000, 1728, 2744, and 4096 atoms, using a lattice constant of 3.52 Å. With such supercells we can carry out a comparison at the same level of accuracy without introducing *a posteriori* "corrections" whose role is to speed the process of finding the extrapolated value of the calculated formation energies.

We report LDA formation energies of a vacancy in the $q = -1$ and $q = -2$ charge states ($V_C^{-1}$ and $V_C^{-2}$) in Figures 2(a) and 2(b). In the case $q = -1$, the calculated values are fitted to quadratic polynomials. The calculated values in both schemes essentially converge to roughly the same value, with the WZP values converging faster. In the case $q = -2$, the calculated values do not appear to converge in either scheme, but the extrapolated values differ by 0.22 eV. Such differences are not negligible because formation energies appear in the exponential for defect concentrations. We also recall that the jellium finite-supercell values, though they appear to be relatively close to the WZP values, they and the corresponding "corrected" values are somewhat ill-defined physically as only their extrapolated values at infinite-supercell size, presumed to be identical[27], define the sought-after formation energy. On the other hand, the calculated WZP values represent the formation energy at a concentration $1/\Omega$ in the limit of no defect pairing or clustering. For example, the formation energy for a 4096-atom supercell, 13.33 eV, corresponding to a concentration of $4 \times 10^{19}$ vacancies per cubic cm, is smaller than the extrapolated value by 0.22 eV.

To examine the role of the neutralizing band state, we calculated the charge density of the defect state and the CBM state in a 512-atom diamond supercell. In Figure 2(c), we show the plane-averaged (dashed line) and the lattice-averaged (solid line) charge densities for the gap state of a neutral vacancy. This plot changes minimally for the two charged states. In Figures 2(d)-2(f) we show a progression of the VBM state for a neutral vacancy and the two charged states. The latter states contain one and two neutralizing holes, respectively. Clearly, the neutralizing VBM states get progressively more localized as a function of charge state, but the localization is significantly less than that of the bound state. Furthermore, the VBM states continue past the defect as Bloch states. Thus, though the supercells are neutral, interactions between the defects persist even for large supercells, causing the variation of the formation energies as functions of supercell size. It is noteworthy that though the VBM is considerably more localized in the case of doubly-charged vacancy, the interactions are stronger because of the double charge. The behavior of formation energies of charged defects in 2D materials is quite different, however, as we show in the next subsection – the differences between jellium and WZP results are substantial.

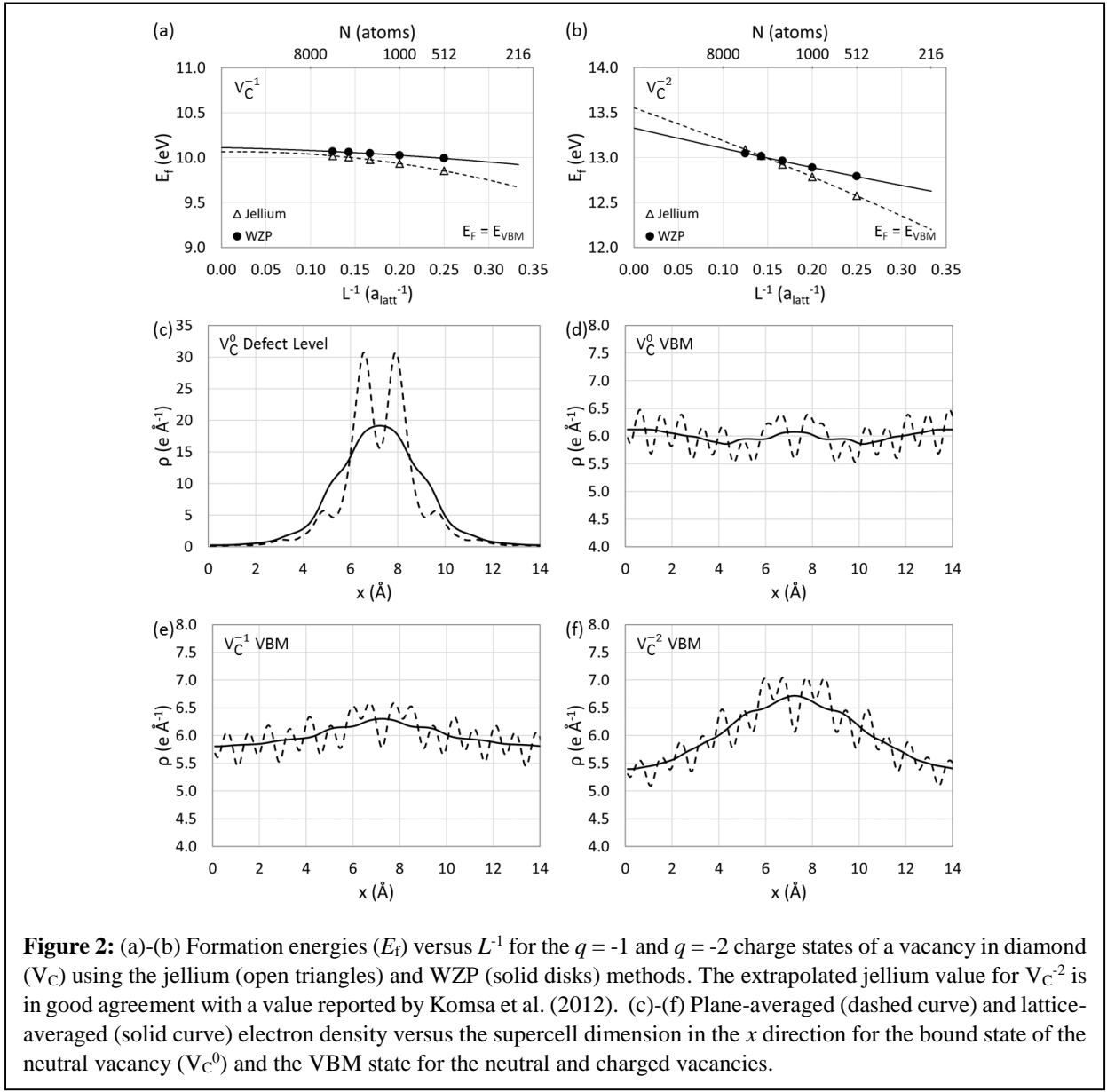

**Figure 2:** (a)-(b) Formation energies ($E_f$) versus $L^{-1}$ for the $q = -1$ and $q = -2$ charge states of a vacancy in diamond ($V_C$) using the jellium (open triangles) and WZP (solid disks) methods. The extrapolated jellium value for $V_C^{-2}$ is in good agreement with a value reported by Komsa et al. (2012). (c)-(f) Plane-averaged (dashed curve) and lattice-averaged (solid curve) electron density versus the supercell dimension in the $x$ direction for the bound state of the neutral vacancy ($V_C^0$) and the VBM state for the neutral and charged vacancies.

We note that the ionized electrons do not occupy bound states, but band states which nevertheless exhibit an amount of localization around the defect, while being Bloch functions at larger distances. This localization feature of band states in the presence of a charged defect [Figures 2(e) and 2(f)] is reminiscent of the continuum wave functions of a hydrogen atom describing unbound electrons in the presence of a proton's Coulomb potential. These wave functions are not plane waves but Whittaker functions,[53, 54] which exhibit a degree of localization around the proton. The localization reflects the effect of the defect's Coulomb potential on the conduction-band states (recall that electron-hole interactions in perfect crystals not only introduce excitons at energies smaller than the band gap, but pull peaks in the continuum interband excitation spectrum down to lower energies as well, especially in wide-gap materials[63, 64]). Even a neutral defect perturbs the band wave functions as seen in Figure 2(d). Indeed, in the case of a hydrogen atom in vacuum, we found that the continuum solutions are essentially plane waves, but that those of lowest energy, analogs of the CBM wave function in Figure 2(d), exhibit minor perturbations in the vicinity of the H atom. In the next subsection we show that the band states are significantly more localized around charged defects in monolayer h-BN.

**IV.B Charged defects in monolayer h-BN**

The results described above for the vacancy in diamond contrast sharply with the results we shall now describe for defects in monolayer h-BN. We use the PBE exchange-correlation functional to calculate formation energies for four charged defects in monolayer BN: the nitrogen vacancy ($V_N^{+1}$, $V_N^{-1}$), a carbon substituting for boron ($C_B^{+1}$) and a carbon substituting for a nitrogen ($C_N^{-1}$). Previous studies[36, 65-67][35, 61-63] have examined the structural and electronic properties of these defects.[16, 18-20] To help us focus on the *formation-energy differences between the two methods*, the atomic chemical potentials are chosen so that the dilute-limit WZP defect formation energy is zero for each defect considered.

We performed calculations for very large crystalline h-BN supercells, ranging from $L$=20 Å ($L^{-1}$ = 0.05 Å$^{-1}$), corresponding to 8x8 unit cells and 128 atoms, to $L$=80 Å ($L^{-1}$ = 0.0125 Å$^{-1}$), corresponding to 32x32 unit cells and 2,048 atoms. The atoms are held fixed in their crystalline positions (no lattice relaxations). The supercell length perpendicular to the plane is set to the length of the in-plane lattice dimension. Supercell lengths are multiples of the primitive lattice constant of 2.5 Å. We performed charged-defect calculations using both the jellium and WZP methods. For the jellium method, once more, we adopt the standard procedure of simply setting the average electrostatic potential to zero [$V(\boldsymbol{G} = 0) = 0$] and invoke no *a posteriori* corrections because the many large supercells allow us to confortably extrapolate to the infinite-supercell limit in both cases in order to compare the difference between the two extrapolated values (recall that *a posteriori* corrections only speed up the process of extracting the infinite-supercell limit, but introduce special uncertainties in the case of 2D materials for which a vacuum region is present; see introduction for details). In the WZP method, we find that the band states occupied by the neutralizing carriers *and the electrostatic potential* are localized to within ±3 Å of the physical h-BN atomic plane, in sharp contrast to the jellium method where the electrostatic potential extends throughout the vacuum regions because of the adoption of a charged defect without its neutralizing band carrier.

In Figure 3, we report results for the unrelaxed $V_N^{+1}$ defect. For the WZP calculations, we adopt a neutralizing band carrier at the CBM, which is appropriate in the case of undoped or *n*-type BN). Figure 3(a) presents the formation energy using the jellium (open triangles) and the WZP (solid disks) methods. We note that *the line through the WZP formation energies is perfectly flat,* with even the small-supercell values being within 0.1 eV of the dilute-limit value. In contrast, the jellium values require a quadratic fit. The WZP result is an indication that neutralization of the supercell by band carriers in BN is very effective in screening the Coulomb tails of the charged defects. The presence of the band carrier is responsible for *the large numerical difference between*

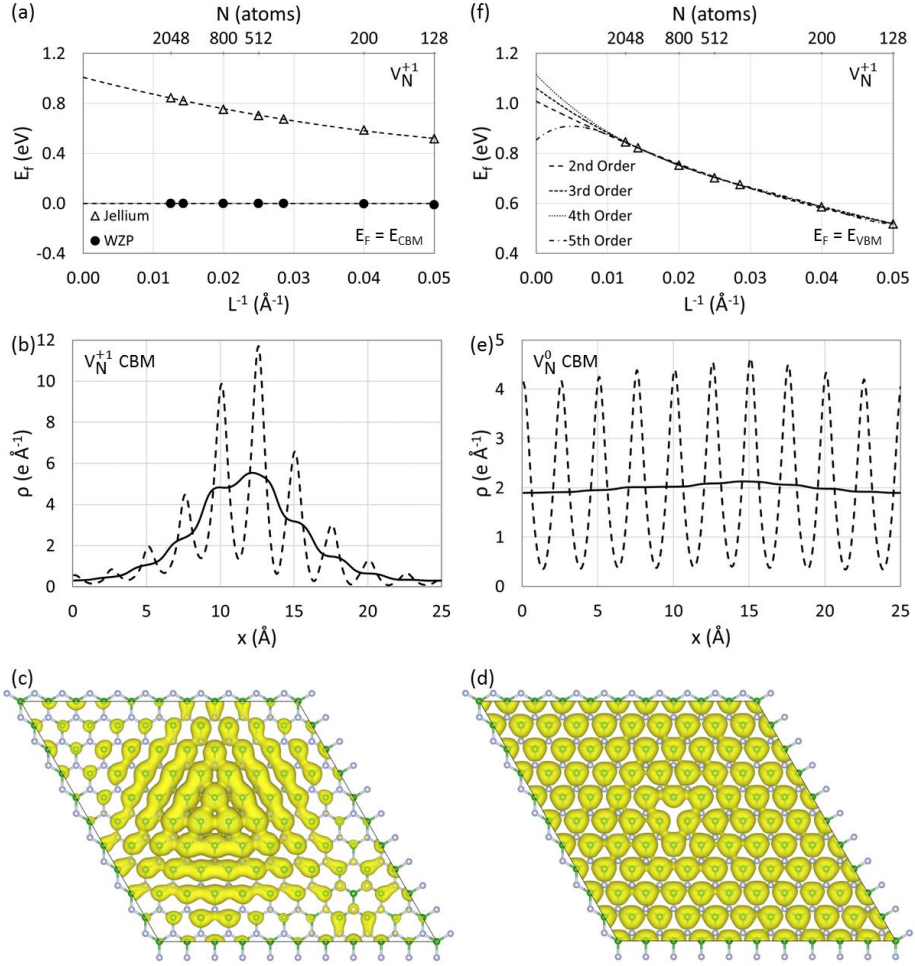

**Figure 3:** (a) Formation energy ($E_f$) versus $L^{-1}$ for the $q = +1$ charge state of the nitrogen vacancy in monolayer h-BN ($V_N^{+1}$) using the jellium (open triangles) and WZP (solid disks) methods with the Fermi level at the VBM. (b) Plane-averaged (dashed curve) and lattice-averaged (solid curve) electron density versus the supercell dimension in the $x$ direction for an occupied CBM state of $V_N^{+1}$ using the WZP method. (c) The $V_N^{+1}$ WZP band-state isosurface ($5 \times 10^{-5}$ $e\text{Å}^{-3}$) superimposed on the ball-and-stick model of the 2D BN supercell showing the localization around the defect and its Bloch nature further out. (d) and (e) Localization of the empty CBM wave function in the presence of a neutral defect. (f) Same panel (a), but the jellium results fitted with high-order polynomials. See text.

*the formation energies obtained by the two methods*, namely the difference in the two extrapolated values, which is 1.0 eV ($\Delta E_f^\infty = 1.0$ eV).

In Figure 3(b), the plane-averaged and lattice-averaged charge densities of the WZP band state are reported for a BN supercell with $L$=25 Å, while in Figure 3(c) we show the same charge density in a 2D plot. The state clearly has a high degree of localization within the supercell with $R_{mm} = 0.05$, though the state is still a Bloch wave function away from the defect. This result differs substantially from the corresponding result for the vacancy in bulk diamond where the band wave function is very delocalized. Further analysis indicates that the localization observed in Figure 3(b) is due to hybridization of the bound defect state with the CBE state, which is consistent with the fact that the neutral $V_N$ defect state derives primarily from conduction bands in hexagonal BN.[62] In contrast, however, the empty CBM state in the case of the neutral vacancy, shown in Figures 3(d) and 3(e) is quite delocalized. ($R_{mm}$ = 0.89). The difference between the $R_{mm}$ values of the CBM states of the neutral and the positively-charged systems indicates that the localization in the latter case reflects the effect of the charged-defect's Coulomb potential on the filled CBM state.

In Figure 3(f) we show the WZP and jellium results one more time. In this plot, however, in addition to the quadratic fit to the jellium results, we show fits with third-order, fourth-order, and fifth-order polynomials. All fit orders have roughly the same coefficient of determination (i.e., $R^2$ value) of >0.99, indicating that the simplest fit (i.e., quadratic) is the most appropriate. The idea that a few DFT-calculated charged-defect formation energies in the jellium approach should be represented by a high-order polynomial[37] or even a high-order polynomial plus an exponential[39] in order to get an upswing in the extrapolation curves at relatively large supercell dimensions was inspired by the fact that *a posteriori* Madelung-like corrections require such high-order fits. Such an upswing, however, was not verified to exist in "uncorrected" DFT-calculated values by doing DFT calculations in large supercells. The results shown in Figure 3(f) demonstrate that no upswing appears for supercell dimensions that exceed the proposed turning point of 16x16 unit cells, which corresponds to $L^{-1}$=0.025 Å$^{-1}$ or 512 atoms in Figure 3(f). The results for supercells ranging up to 32x32 unit cells or $L^{-1}$=0.0125 Å$^{-1}$ (2048 atoms) can be fitted just as accurately with a quadratic. The net conclusion here is that there is considerable uncertainty in how to implement *a posteriori* corrections in the case of charged defects in 2D materials and perform extrapolations to the infinite-supercell limit in the jellium scheme. There are no such uncertainties in the WZP approach.

The hybridization of the conduction-band states with the bound states and concomitant localization of the former around the defect persists beyond the CBM state, causing different degrees of localization in more than ten states. In Figures 4(a) and 4(b), we display the state density of the CBM and the 6$^{th}$ higher-energy conduction band, respectively, in the presence of the $V_N^{+1}$ defect. The same state densities are shown in Figures 4(c) and 4(d) at a smaller isosurface level. This persistent localization of band states that far from the defect are Bloch states is an even more vivid demonstration of the effect of the charged-defect potential on band states. As we noted in the case of charged vacancies in diamond [Figures 2(e) and 2(f) in Section IV.A], such localization is an analog of what happens to the continuum states of a hydrogen atom, which are described by Whittaker functions[53, 54]. We recalculated the formation energy for $V_N^{+1}$ by placing the neutralizing electron in each of the thirteen lowest-energy conduction bands and found that the formation energy increases in the range 0.1 to 0.3 eV. Thus, an average over the Fermi-Dirac distribution function would result in a temperature-dependent formation energy.

Figure 5 reports our results for the unrelaxed $V_N^{-1}$ defect in monolayer BN. In this case, for WZP calculations, we adopt a hole at the VBM as the neutralizing carrier, which is appropriate for undoped or *p*-type BN. The formation energies versus inverse supercell size are reported in Figure 5(a), with the Fermi level $E_F$ at the VBE. The WZP curve is again completely flat. In contrast, the jellium curve has a considerable slope and requires a quadratic fit. Once more, no upswing is observed at supercell dimensions larger than $L$=40 Å ($L^{-1}$=0.025 Å$^{-1}$), confirming a persisting uncertainty as to how to implement *a posteriori* corrections and extrapolate DFT-calculated charged-defect formation energies in the jellium scheme.

For this defect, we find a less pronounced but still substantial difference between the formation energies obtained using the jellium and WZP approaches. The dilute-limit formation energies differ by 0.52 eV. As shown in Figures 5(b) and 5(c), the WZP neutralizing state, unlike the case of vacancies in bulk diamond discussed above, exhibits substantial localization about the defect, with $R_{mm} = 0.24$. Further analysis of the results indicates that the localization observed in Figure 5(b) for $V_N^-$ is smaller than the localization observed in Figure 3(a) for $V_N^{+1}$ because of less hybridization of the defect state in the gap with the VBM state than with the CBM state.

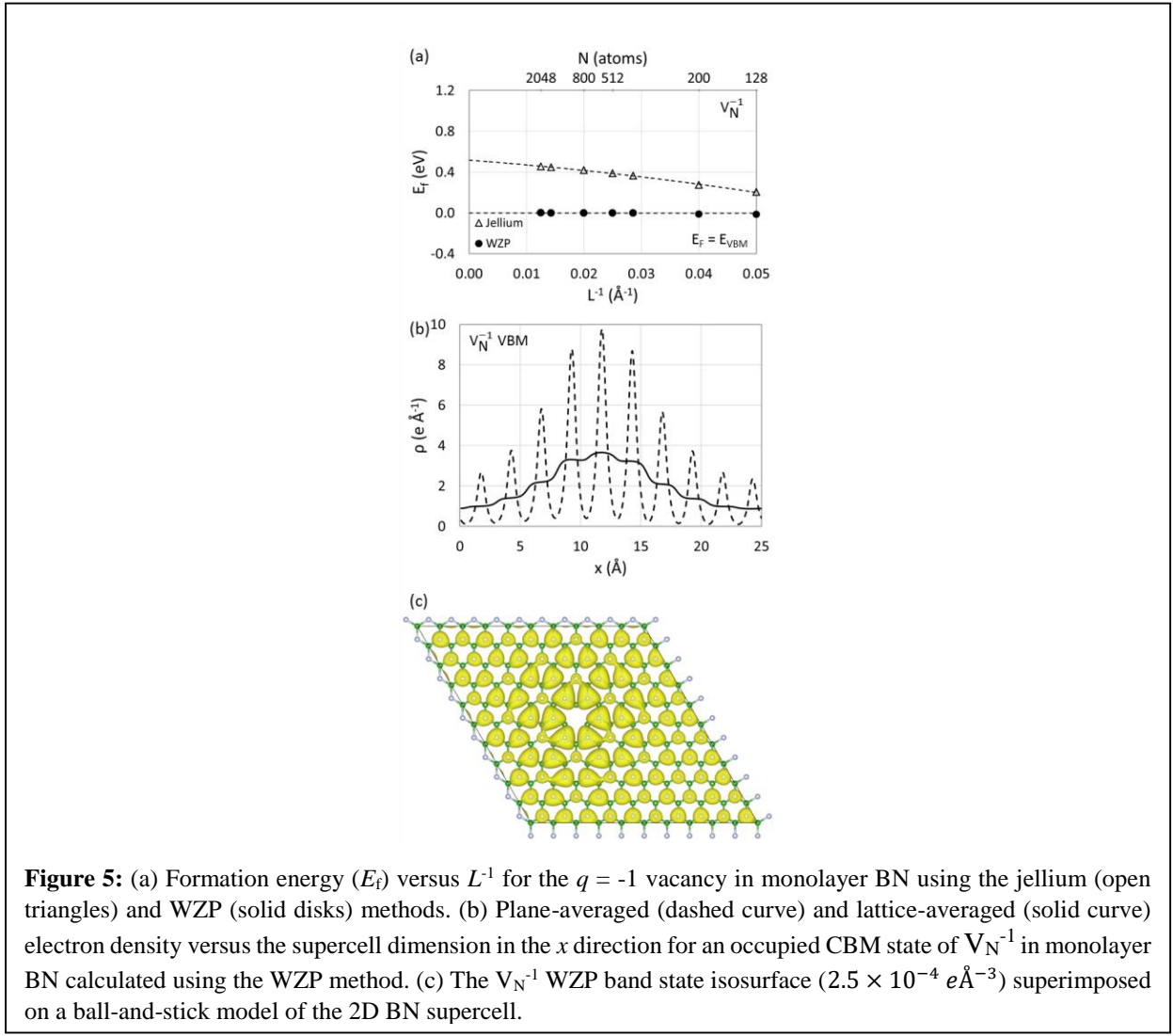

**Figure 5:** (a) Formation energy ($E_f$) versus $L^{-1}$ for the $q$ = -1 vacancy in monolayer BN using the jellium (open triangles) and WZP (solid disks) methods. (b) Plane-averaged (dashed curve) and lattice-averaged (solid curve) electron density versus the supercell dimension in the *x* direction for an occupied CBM state of $V_N^{-1}$ in monolayer BN calculated using the WZP method. (c) The $V_N^{-1}$ WZP band state isosurface ($2.5 \times 10^{-4}$ $e\text{Å}^{-3}$) superimposed on a ball-and-stick model of the 2D BN supercell.

Figure 6 reports our results for the unrelaxed $C_B^{+1}$ defect in monolayer BN. The formation energies calculated with the Fermi level chosen at the CBM are reported in Figure 6(a). The WZP formation-energy curve is flat as in the case of $V_N^{+1}$, whereas the jellium formation energy varies linearly with inverse lattice length. In the dilute-limit, the jellium result is larger than the WZP result, with $\Delta E_f^\infty = 1.03$ eV. In Figure 6(b), the averaged densities of the WZP band state are reported. The state is clearly highly localized, with $R_{mm} = 0.07$. As is the case for the empty CBM state in the presence of a neutral $V_N$ defect, the empty CBM state in the presence of a neutral $C_B$ defect is again highly delocalized, as shown in Figures 6(d) and 6(e), with $R_{mm} = 0.62$.

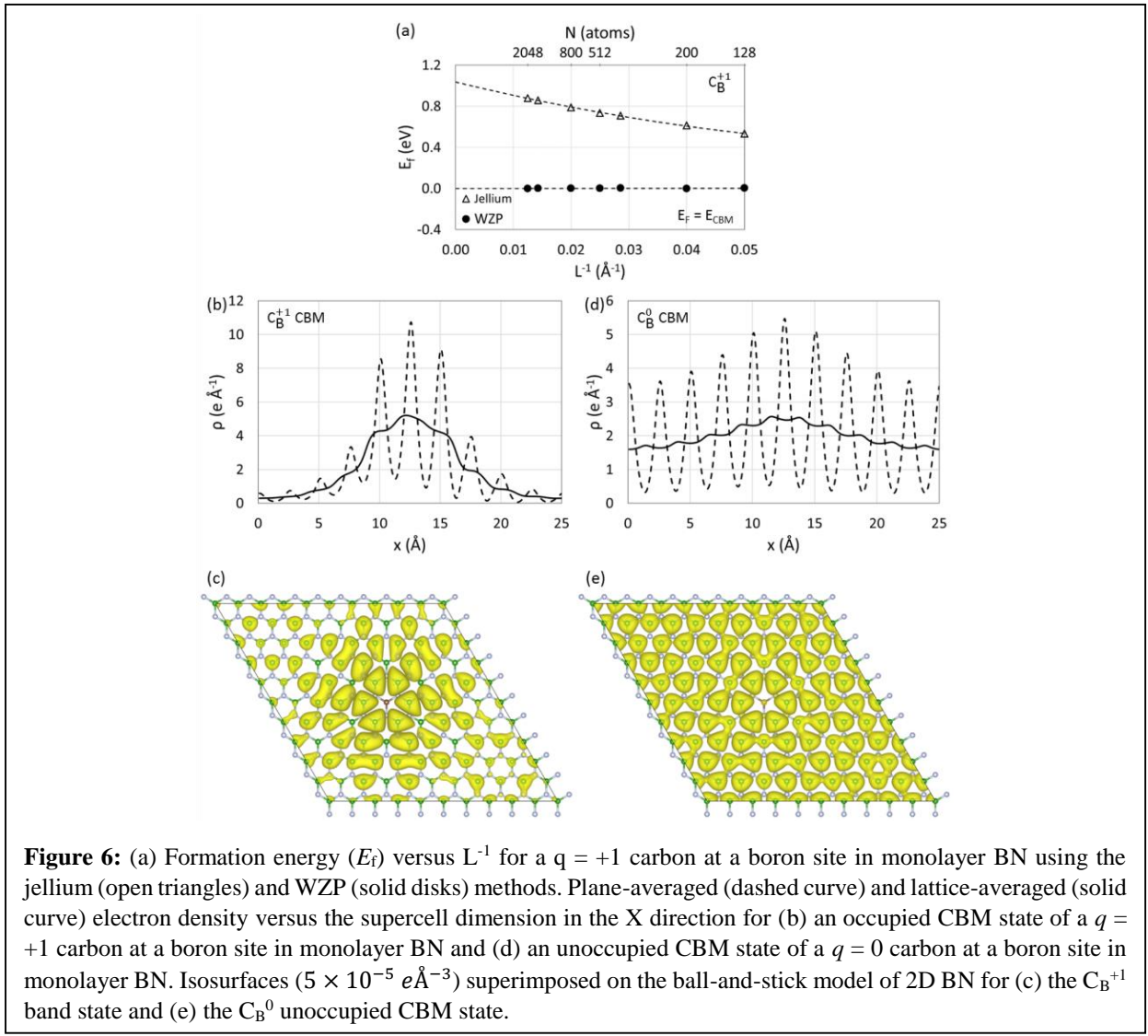

**Figure 6:** (a) Formation energy ($E_f$) versus $L^{-1}$ for a q = +1 carbon at a boron site in monolayer BN using the jellium (open triangles) and WZP (solid disks) methods. Plane-averaged (dashed curve) and lattice-averaged (solid curve) electron density versus the supercell dimension in the X direction for (b) an occupied CBM state of a $q = +1$ carbon at a boron site in monolayer BN and (d) an unoccupied CBM state of a $q = 0$ carbon at a boron site in monolayer BN. Isosurfaces ($5 \times 10^{-5}$ $e\text{Å}^{-3}$) superimposed on the ball-and-stick model of 2D BN for (c) the $C_B^{+1}$ band state and (e) the $C_B^0$ unoccupied CBM state.

In Figure 7(a), formation energies are reported for the unrelaxed $C_N^{-1}$ defect with the Fermi level at the VBM. The WZP curve is again completely flat. The jellium values, on the other hand, show a considerable slope and can be fitted well with a quadratic, without a sharp upswing at very large supercells. The infinite-supercell extrapolated values from the two different methods differ by 1.01 eV, i.e., $\Delta E_f^\infty = 1.01\ eV$. In Figure 7(b), the averaged densities of the WZP band state are presented. The WZP VBE neutralization state is less localized for the $C_N^{-1}$ defect than for the $V_N^{+1}$ defect but more localized than the $V_N^{-1}$ defect. The averaged density has a localization in between the extremes observed for the previous defects, with $R_{mm} = 0.11$.

The above calculations for charged defects in monolayer BN demonstrate clearly that, in addition to its robust statistical-mechanics-based derivation, the WZP approach is superior to the jellium approach *for practical calculations as well*. For all the charged defects we investigated in monolayer BN, the WZP DFT-calculated formation energies converge quickly using relatively small supercells containing ~200 atoms and remains completely flat. The calculations are straightforward without having to deal with elimination of divergences, a posteriori "corrections" to remove long-range interactions, and extrapolations that may or may not have upswings or downswings after a certain, fairly large supercell dimension.

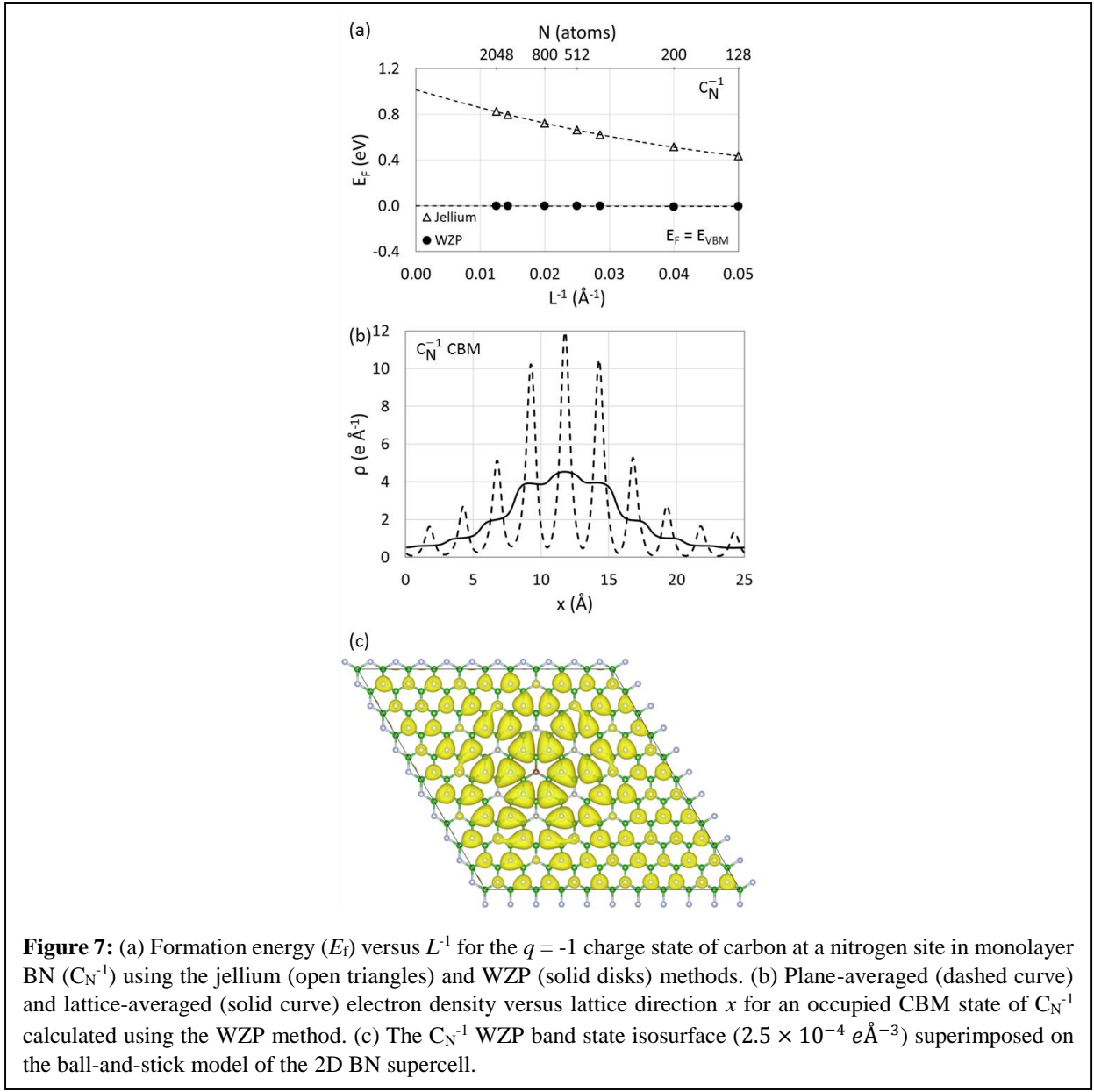

**Figure 7:** (a) Formation energy ($E_f$) versus $L^{-1}$ for the $q = -1$ charge state of carbon at a nitrogen site in monolayer BN ($C_N^{-1}$) using the jellium (open triangles) and WZP (solid disks) methods. (b) Plane-averaged (dashed curve) and lattice-averaged (solid curve) electron density versus lattice direction $x$ for an occupied CBM state of $C_N^{-1}$ calculated using the WZP method. (c) The $C_N^{-1}$ WZP band state isosurface ($2.5 \times 10^{-4}\ e\text{Å}^{-3}$) superimposed on the ball-and-stick model of the 2D BN supercell.

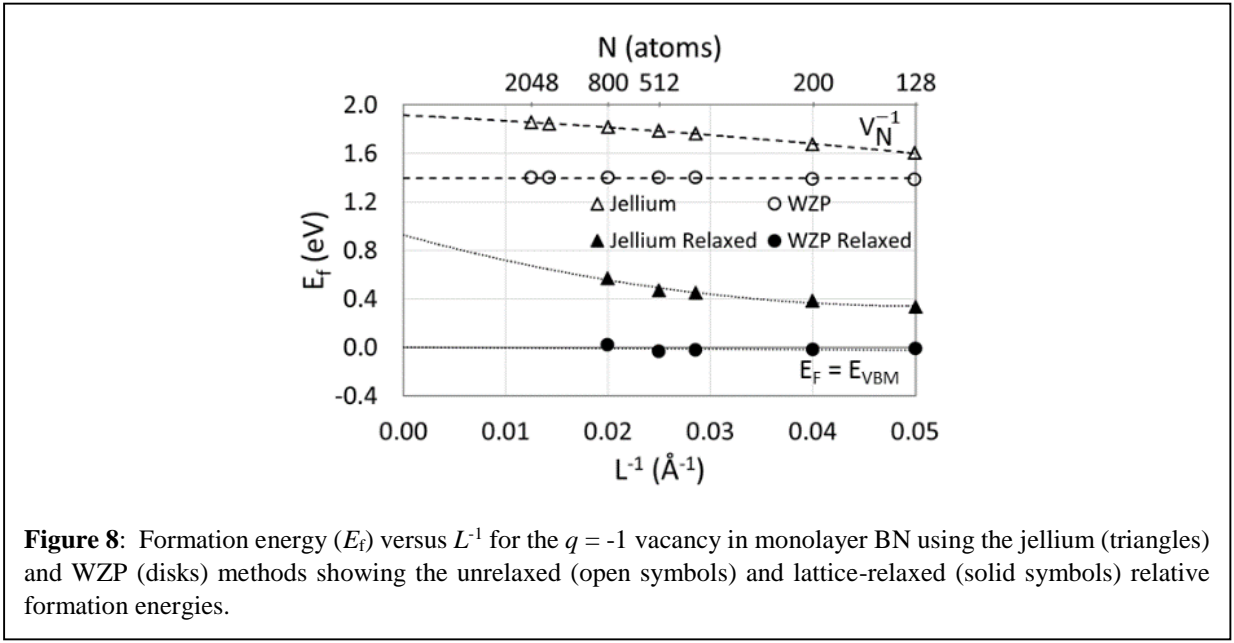

**Figure 8**: Formation energy ($E_f$) versus $L^{-1}$ for the $q = -1$ vacancy in monolayer BN using the jellium (triangles) and WZP (disks) methods showing the unrelaxed (open symbols) and lattice-relaxed (solid symbols) relative formation energies.

All the calculations reported so far for charged defects in BN are for unrelaxed configurations, which enabled us to perform calculations for very large supercells and compare the two methods as directly as possible. We have performed calculations with full lattice relaxation for one defect, namely $V_N^{-1}$. The results are shown in Figure 8. It is clear that relaxations have a large effect on formation energies. The faster convergence of the WZP results is again evident, as about 200 atoms are sufficient (in fact larger supercells entail larger numerical errors), whereas the jellium results need to be extrapolated.

### IV.C Quantitative comparison of WZP and jellium features

Having compared the formation energies of charged defects in monolayer h-BN, in this Section we provide additional insights into the differences between the jellium and WZP approaches by displaying and discussing other features in a quantitative way. First, we show that the WZP approach does not suffer from a problem that occurs in the jellium method when applied to charged defects at surfaces and 2D materials – the jellium-based formation energies diverge if the vacuum region is increased while keeping the in-plane supercell dimension fixed[13, 14, 35]. The divergent feature is demonstrated by the open triangles in Figure 9(a) for the $C_N^{-1}$ defect. The divergence can be avoided by "scaling" the three supercell dimensions, but the divergence remains latent with no possibility to investigate potential consequences. Corresponding formation energies obtained using the WZP method, shown as dark circles in Figure 9(a), exhibit no such divergence but remain converged. This difficulty of the jellium method in 2D materials is in addition to uncertainties in extrapolating either "uncorrected" or "corrected' formation energies to the infinite-supercell limit, as discussed earlier in this paper.

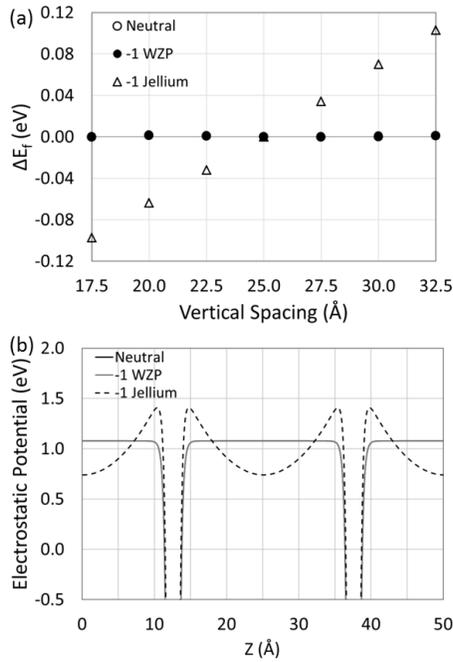

**Figure 9:** (a) Change in formation energy as a function of vacuum spacing for the $C_N$ defect in the neutral state and the $q = -1$ charge state using both the WZP and jellium methodologies. The open and black circles overlap fully. (b) Planar averaged electrostatic potential along the z-axis. The WZP potential of the $q = -1$ charge state is indistinguishable from the potential of the neutral state.

In Figure 9(b), we examine the averaged local electrostatic potential for the neutral $C_N^0$ and the charged $C_N^-$ defects in the WZP and jellium approaches. The figure shows the planar average of the local electrostatic potential along the z-axis (i.e. perpendicular to the BN plane) in a 25-Å supercell. *The local potential of the q=-1 charge state using the WZP methodology is indistinguishable from that of the charge-neutral state* on the scale of Figure 9(b) – it is totally flat in the vacuum region, *demonstrating the intrinsic absence of any unphysical long-range potentials*. In sharp contrast, in the case of the *q*=-1 charge state calculated using the jellium scheme, there exist long-range potentials in the vacuum regions that persist even in very large supercells. These potentials arise because of the adopted definition of a charged defect as a defect for which one or more electrons have been removed altogether from the crystal or added into the crystal, instead of trading electrons with band states, which is the physical reality. Note that these long-range potentials persist in the vacuum regions even after the long-range *interactions* generated by these potentials are removed by an *a posteriori* "correction" and in the infinite-supercell limit. Indeed, the central claim of the jellium approach is that, in the infinite-supercell limit, the long-range potential emanating from charged defects is purely Coulombic (unscreened Coulombic in the vacuum region of supercells for 2D materials and surfaces).

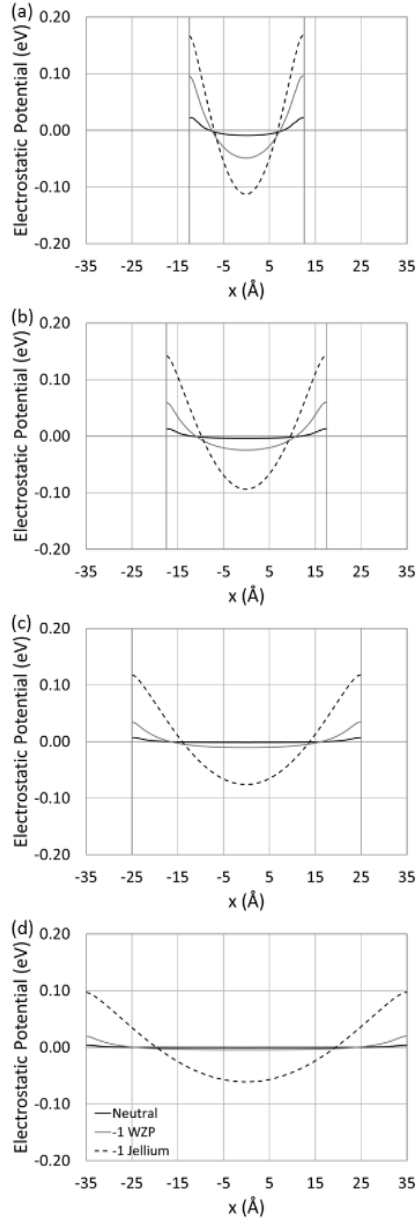

**Figure 10:** Evolution of the planar-averaged electrostatic potential along the *x*-axis for the $C_N$ defect in the neutral state and the *q* = -1 charge state using both the WZP and jellium methodologies utilizing (a) 10x10, (b) 14x14, (c) 20x20, and (d) 28x28 supercells. While the jellium potential remains quadratic in each supercell size, the WZP potential becomes flatter and more like the neutral state as the supercell size increases.

In Figure 10, we show the averaged electrostatic potentials as they *evolve in the in-plane supercells* in the case of the $C_N^{-1}$ defect. The zero point on the x-axis is the mid-point between two adjacent defects, which reside at the positions of the vertical lines. It is clear that, even in the neutral-defect case, there is an electrostatic interaction between defects that vanishes quickly for large supercells. For charged defects in the WZP scheme, the electrostatic interaction is also very small and vanishes quickly because of the localization of the band states that neutralize the defects.

The electrostatic interaction between charged defects in the jellium scheme, however, is substantial and persists as the supercells get large because the defects are truly charged and the supercells are not neutralized by a hypothetical neutralizing uniform charge density (the curves have a parabolic form). These interactions and those through the potentials in the vacuum regions are responsible for the persistent slope of the curves fitted to jellium formation energies, causing difficulties with obtaining accurate extrapolations to infinite-supercells.

We now address the validity of the claim that, in the jellium scheme, setting $V(\boldsymbol{G}=0)=0$ is *equivalent* to introducing a "homogeneous background charge density" because such a charge density, which is *not* added to the DFT electron density $\rho^+(r)$, affects only the $\boldsymbol{G}=0$ term of the defect potential, "neutralizes" the supercells as far as the defect potential is concerned, and thus removes the divergence. Clearly, the claimed equivalence is true if and only if a homogeneous charge density affects only the $\boldsymbol{G}=0$ term of the potential. To our knowledge, though this assertion is made explicitly in the original 1984 paper using supercells for total-energy calculations[10] and elsewhere, it has never been proved.

Let us, therefore, examine the potential that is generated by a homogeneous charge density $\rho_h(r) = constant$. Taking symmetry into account, the solution of Poisson's equation is $v(\boldsymbol{r}) = Ar^2 + B$, where $A$ is fixed by the value of the constant and $B$ has no significance as it can set the zero of energy. This solution is valid for any coordinate origin $\boldsymbol{r}=0$. If we impose the periodicity of our supercells, e.g., monolayers with vacuum regions, the solution is valid if the origin is at the mid-point between monolayers along a line connecting equivalent defects. One can easily verify that the Fourier transform of $v(\boldsymbol{r}) = Ar^2$ in an infinite space, whether homogeneous or carved into periodic supercells, is such that $v(\boldsymbol{G}=0)$ is infinite. However, *the other Fourier components, $v(\boldsymbol{G} \neq 0)$, are not zero*. In other words, setting $V(\boldsymbol{G}=0)=0$ is not equivalent to introducing a "homogeneous background charge density" because such a charge density would alter the $V(\boldsymbol{G} \neq 0)$ components of the defect potential. The net conclusion is that setting $V(\boldsymbol{G}=0)=0$ removes the divergence simply because the divergent average defect potential, $V(\boldsymbol{G}=0)$ is set to zero. The $V(\boldsymbol{G} \neq 0)$ components of the defect potential remain intact, which means that the overlapping Coulombic tails that result from $\rho^+(\boldsymbol{r})$ in the DFT calculation are not altered in any way, i.e., *the supercells are not "neutralized"*. Overall, the moniker "jellium" is inappropriate for the method as no kind of jellium is ever introduced into the theory.

The above findings suggest that the electrostatic potential in the vacuum region of the jellium supercells, as seen in Figure 9(b), which can be fit to a parabola centered at the midpoint between defects in neighboring supercells, has nothing to do with a hypothetical background homogeneous charge density, which has been the central tenet of the jellium approach. Such parabolic shapes are also present along the in-plane line connecting neighboring charged defects (see Figure 10) and in the jellium defect potential of charged defects in diamond that we described earlier in this paper (not shown). They have also been found in the case of jellium calculations for charged defects in bulk materials in prior work[21, 23]. It is now clear that these parabolas arise strictly from the overlapping Coulomb tails.

We can illustrate the above conclusions with a simple model calculation of superposed Coulombic tails. We took a one-dimensional array of $N$ point charges and calculated the electrostatic potential in the region between the middle pair of charges. As a function of $N$, this potential quickly converges to essentially a perfect parabola of the form $V(x) = Ax^2 + B(N)$. The constant $A$ remains unchanged as $N$ is increased, but $B(N)$ increases slowly, but monotonically, i.e., it diverges as $N \to \infty$. The parabolas that have been found by DFT calculations suggest that a three-dimensional model behaves in a similar way. The net conclusion is that the parabolas of jellium calculations in Figures 8(b) and 9 arise from overlapping Coulomb tails, not from a hypothetical homogeneous background charge density.

We can gain additional insights into the so-called "jellium" scheme as follows. If we sit at the mid-point between two defects in adjacent supercells, the Coulombic tails from the infinite periodic array of charged defects (one in each supercell) arrive at that point. $V(\boldsymbol{G} = 0)$, the average electrostatic potential, is infinite simply because we assume an infinite number of supercells. As usual, when we take an infinite limit, it matters how we do it. If we were to say that the number of supercells is not really infinite but an extremely large number, the average electrostatic potential would just be an extremely large constant value that does no harm because it is a constant and can be set to zero. The Coulombic tails are still there because all the other Fourier components of the defect potential have remained untouched. We can take the limit to infinity at that point, keeping $V(\boldsymbol{G} = 0) = 0$, while retaining all Coulombic tails intact. The supercells are not really neutralized as no charge of any kind is ever added. Indeed, at every iteration of the "jellium" self-consistency loop, the average electrostatic potential $V(\boldsymbol{G} = 0)$ is automatically set to zero by DFT computer codes, without doing anything to the non-constant part of the overlapping Coulombic tails themselves [the $V(\boldsymbol{G} \neq 0)$ terms], i.e., the supercells are not neutralized. The parabolas we mentioned earlier in the "jellium" defect potential of Fig. 9b, in Figure 10, and in jellium calculations of charged defects in bulk semiconductors are simply the result of superposing these Coulombic tails.

The above analysis actually puts the "jellium" approach on a firmer footing, at least for charged defects in bulk materials. The jellium approach essentially treats charged supercells as in Figure 1(d), which are never neutralized in any way. If the total number of periodic supercells is viewed as extremely large but not quite infinite, $V(\boldsymbol{G} = 0)$ is finite and can be set to zero. *Poisson's equation is not violated* and the method yields legitimate formation energies for truly charged, not merely ionized, defects at a concentration corresponding to the supercell size in the limit of no defect pairing or clustering. In the infinite-supercell limit, the result would truly match the results of Green's-function calculations, as that was done in the 1980's[6-8]. The only problem is that charged defects in physical crystals are not truly charged, i.e., electrons are not removed from or added to the crystal, but they are merely traded with the energy bands, always leaving the crystal neutral. We can now understand why for the charged defects in diamond that we studied in Section IV.A the jellium and WZP formation energies are so similar, especially for $q = -1$. The only difference is inclusion of the band carrier in the calculation of the electron density in the method. In the absence of significant localization of the band wave function about the defect, the effect is rather small. For the $q = -2$ case, the band state is more localized around the defect, but the larger $q$ value enhances the difference made by the neutralizing band electron. The point remains that one cannot know ahead of time how important is the effect, *whereby the only safe way to proceed is to adopt the WZP method because the neutralizing band carrier is de facto there.*

The case of charged defects at surfaces and in 2D materials, on the other hand, renders the choice of WZP as the method of choice more directly. Even though the jellium method does not really invoke a hypothetical uniform charge density in the vacuum region that affects the potential, electrostatic potentials are indeed present in the vacuum regions, as shown in Figure 9(b), and are totally unphysical as they arise from a supercharged surface or 2D material (they are also very strong because they are totally unscreened in the vacuum regions). In contrast, *the physical surface or 2D material are always neutral, with their potentials totally confined within the regions occupied by the material.* Thus, performing a calculation that entails unphysical potentials in the vacuum regions and then invoking an *a posteriori* correction to remove the long-range defect interactions (but not the long-range potentials themselves), even if numerically successful, yields formation energies for truly "charged" defects as in the schematic of Figure 1(d), which do not correspond to physical reality. The large differences in both formation energies and convergence rates of the two methods for charged defects in h-BN and MoS$_2$ given in this paper make the case for the WZP approach.

**IV.C Charged defects in other 2D systems**

In addition to BN, which has a wide band gap, we performed one test for a defect in MoS$_2$, which has a smaller band gap and is a true semiconductor. Formation energies for an unrelaxed negatively charged sulfur vacancy are shown in Figure 11. The curve for WZP values is again completely flat, showing very fast convergence, whereas the jellium results require a quadratic fit to get an extrapolated value. The two obtained formation-energy values now differ by only 0.21, which, though smaller than the differences we found for defects in BN, it is still substantial, given that such formation energies typically occur in exponentials. Note, however, that for jellium we report results for supercells up to 1200 atoms (20x20 unit cells) and find that *there is no evidence for an up- or down-swing requiring a 5$^{th}$-order polynomial fit in the calculated formation-energy values at such large supercells* as inferred in recent papers[37, 39] from the corresponding behavior of *a posteriori* corrections.

We finally address some peculiar issues that arise in the case of defects in vertical heterostructures between two 2D materials. Due to the periodic boundary condition along the direction perpendicular to the surface, systems with interfaces or surface adsorbates develop an artificial electric field through the vacuum. Most DFT codes now include corrections to the total energy, local potential, and forces to remedy the problem[68, 69]. Such correction schemes, however, often have issues when charged defects are introduced. The VASP code, for example, clearly warns that potential and force corrections are only implemented for charged cells with cubic geometry – a restriction that cannot easily accommodate large-scale interface slab calculations with defects. As the WZP supercell always remains truly neutral, the WZP methodology is inherently compatible with the standard dipole correction techniques. As an example, we present an unrelaxed calculation using an MoS$_2$/BN interface constructed with an $8 \times 8$ supercell of the MoS$_2$ primitive cell and a strained $10 \times 10$ BN supercell (lateral size 25.32 Å) and include a sulfur vacancy (V$_S$). The vacuum spacing between periodic images was set equal to the lateral spacing. In this case, the dipole-induced electric field is $1.14 \times 10^{-11}$ V/m. As this is a relatively small field, the inclusion of the dipole correction terms only changes the total energies by ~2 meV for the defect-free interface, neutral vacancy, and WZP negatively charged vacancy ($V_S^{-1}$). While VASP gives an explicit error message for the inclusion of local potential corrections and force corrections in the jellium case, no such warning is given if only the total-energy correction flag is chosen. In this case, the energy correction adds 350.66 eV to the total energy – a clearly erroneous amount. Under the WZP methodology, the 0/-1 charge transition level (i.e. the difference in formation energy at a fixed Fermi energy) is 1.26 eV for the sulfur vacancy at the interface. Comparatively, the charge transition level is 1.28 eV for a free monolayer in the WZP methodology, indicating a small effect in the present case due to the interface. The level for the jellium model, by contrast, is 1.49 eV in the interface-free MoS$_2$ monolayer. While these effects are small in the present example, interfaces with more complicated bonding or interfaces with intercalated atoms and molecules would be expected to have larger interface dipoles and potentially different shifts in defect formation energies and transition levels. The WZP methodology, on the other hand, allows for charged defects to be studied in a straightforward manner in such situations.

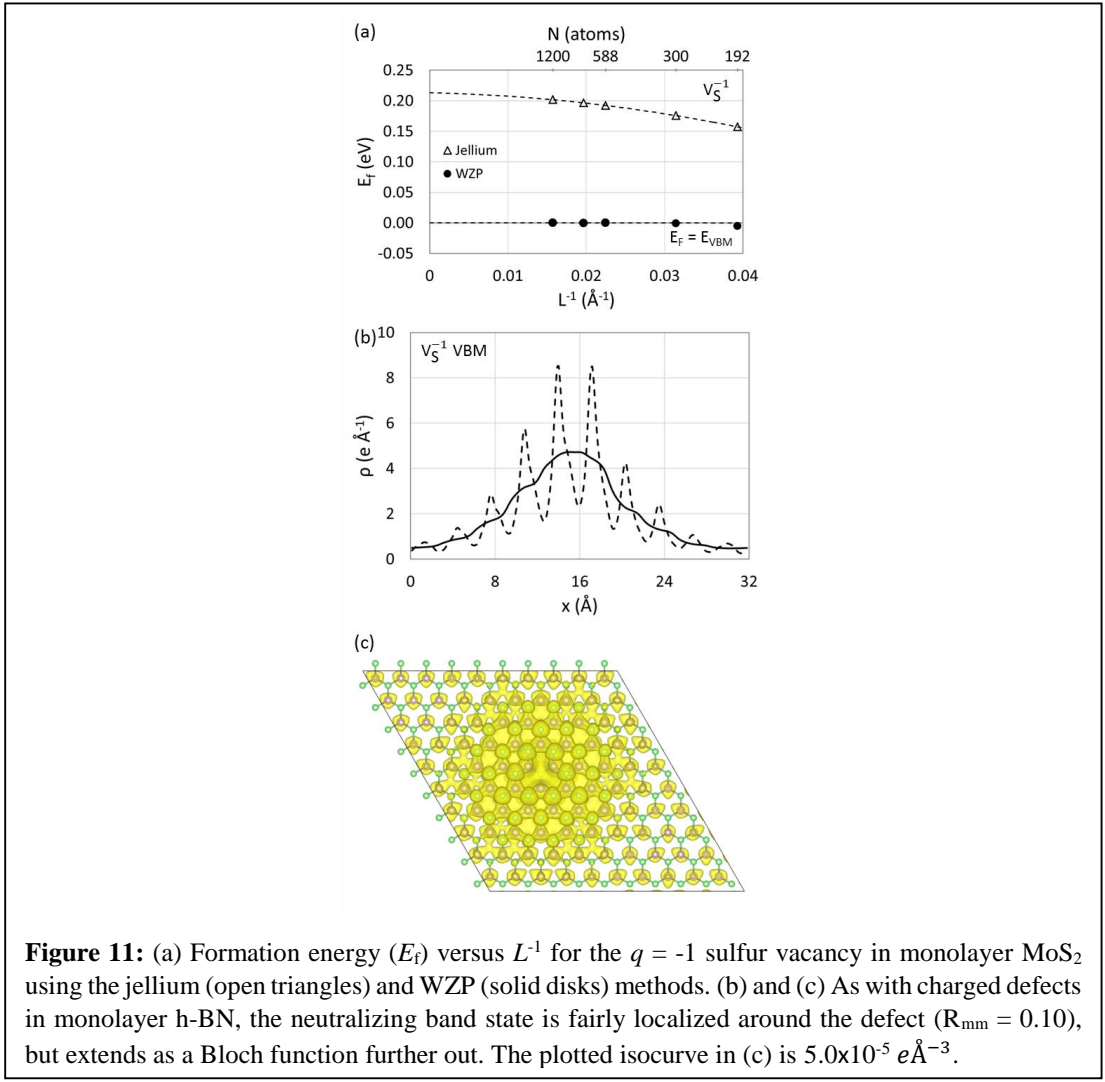

**Figure 11:** (a) Formation energy ($E_f$) versus $L^{-1}$ for the $q = -1$ sulfur vacancy in monolayer MoS$_2$ using the jellium (open triangles) and WZP (solid disks) methods. (b) and (c) As with charged defects in monolayer h-BN, the neutralizing band state is fairly localized around the defect ($R_{mm} = 0.10$), but extends as a Bloch function further out. The plotted isocurve in (c) is 5.0x10$^{-5}$ $e\text{Å}^{-3}$.

V. **Summary**

In this paper we gave an extensive discussion of the evolution of the conventional supercell approach to calculations of formation energies of charged defects, highlighting the complications that arise from the presence of a divergence and long-range Coulomb interactions and the ways that have been used to deal with them, especially the so-called jellium method. We then gave a brief summary of the recent WZP paper, where the case was made that the divergence and all other complications arise from the fact that the conventional approaches view "charged" defects as truly charged, i.e., charged defects are formed by removing electrons from or adding electrons to the crystal, leaving them with screened Coulomb tails that extend all the way to infinity. This scenario is not in accord with physical reality and the principles of statistical mechanics. "Charged" defects are in fact only ionized, i.e., they become charged by trading electrons and holes with the energy bands or dopant states, so that the crystal always remains neutral. Indeed, the Fermi level in semiconductors is obtained from the charge neutrality condition. The WZP reformulated the theory of charged-defect formation energies by recognizing that for each defect, there are a corresponding number of electrons in the conduction bands or holes in the valence bands so that every supercell contains just the right number of band carriers to neutralize the supercell. The new definition does not lead to any divergences or other complications that need to be resolved by ad hoc procedures.

The present paper presents the WZP theory in great detail and shows that the jellium definition of charged-defect formation energies can be derived from the statistical-mechanics-backed WZP definition by invoking a series of approximations whose validity cannot be assessed *a priori* (uncontrolled approximations). A number of calculations have been presented for charged defects in both bulk crystals and 2D materials, which demonstrate that the two theories yield significantly different numerical results, establishing the superiority of the WZP theory as a benchmark. Analysis of the defect potentials led to deep insights in the workings of both the WZP and the jellium methods.

**Acknowledgements -** This research was funded by NSF grant RUI-DMR 1506403 (BRT) and by NSF grant ECCS-1508898 at Vanderbilt University and the University of Florida. At Vanderbilt University, the work was further supported by DTRA grant HDTRA1-16-1-0032, DOE grant DE-FG02-09ER46554 and by the McMinn Endowment. Calculations were performed using Advanced CyberInfrastructure computational resources provided by The Institute for CyberScience at The Pennsylvania State University (http://ics.psu.edu) and the Department of Defense's High Performance Computing Modernization Program. The authors declare that there is no conflict of interest regarding the publication of this paper.